\documentclass{jfm}
\usepackage{graphicx}
\usepackage{epstopdf, epsfig}
\usepackage{color}

\shorttitle{Jump rope vortex flow in liquid metal Rayleigh-Bénard convection}
\shortauthor{M.Akashi, T.Yanagisawa, A.Sakuraba, F.Schindler, S.Horn, T.Vogt, and S.Eckert}

\title{Jump rope vortex flow in liquid metal Rayleigh-Bénard convection in a cuboid container of aspect ratio five}

\author{{Megumi Akashi}\aff{1}
\corresp{\email{m.akashi@hzdr.de}}, Takatoshi Yanagisawa\aff{2}, Ataru Sakuraba\aff{3}, Felix Schindler\aff{1}, Susanne Horn\aff{4}, Tobias Vogt\aff{1} \and Sven Eckert\aff{1}}

\affiliation{\aff{1}Department of Magnetohydrodynamics, Institute of Fluid Dynamics, Helmholtz-Zentrum Dresden-Rossendorf, Bautzner Landstraße 400, D-01328 Dresden, Germany
\aff{2}Japan Agency for Marine-Earth Science and Technology (JAMSTEC), 2-15 Natsushima-cho, Yokosuka-city, Kanagawa, 237-0061, Japan
\aff{3}Department of Earth and Planetary Science, Graduate School of Science, The University of Tokyo, 7-3-1 Hongo, Bunkyo-ku, Tokyo, 113-0033, Japan
\aff{4}Centre for Fluid and Complex Systems, Coventry University, Mile Lane, CV1 2NL, Coventry, United Kingdom}

\begin{document}

\maketitle

\begin{abstract}

We study the topology and the temporal dynamics of turbulent Rayleigh-Bénard convection in a liquid metal with a Prandtl number of 0.03 located inside a box with a square base area and an aspect ratio of $\Gamma = 5$. Experiments and numerical simulations are focused on the Rayleigh number range, $6.7 \times 10^4 \leq Ra \leq 3.5 \times 10^5$, where a new cellular flow regime has been reported by a previous study (Akashi \textit{et al.}, \textit{Phys. Rev. Fluids}, vol.4, 2019, 033501). This flow structure shows symmetries with respect to the vertical planes crossing at the center of the container. The dynamic behaviour is dominated by strong three-dimensional oscillations with a period length that corresponds to the turnover time. Our analysis reveals that the flow structure in the $\Gamma$ = 5 box corresponds in key features to the jump rope vortex structure, which has recently been discovered in a $\Gamma = 2$ cylinder (Vogt \textit{et al.}, \textit{Proc. Natl Acad. Sci. USA}, vol.115, 2018, 12674-12679). While in the $\Gamma = 2$ cylinder a single jump rope vortex occurs, the coexistence of four recirculating swirls is detected in this study. 
Their approach to the lid or the bottom of the convection box causes a temporal deceleration of both the horizontal velocity at the respective boundary and the vertical velocity in the bulk, which in turn is reflected in Nusselt number oscillations. The cellular flow regime shows remarkable similarities to properties commonly attributed to turbulent superstructures.

\end{abstract}

\begin{keywords}
Rayleigh-Bénard convection, thermal turbulence, heat transport, large-scale circulation, liquid metal, low Prandtl number
\end{keywords}

\section{Introduction}
Rayleigh-Bénard convection (RBC) is a classical problem in fluid dynamics that has been studied for more than 100 years and serves as a model system for thermally-driven flows in nature and engineering \citep{Benard1900, Chandrasekhar1961, Ahlers2009, Chilla2012}. The RBC configuration considers a fluid exposed to a destabilizing vertical temperature gradient between two plane-parallel plates and is described by the Rayleigh number, the Prandtl number and the aspect ratio of the fluid volume: 
\begin{equation}
Ra = \frac{\alpha{g}\Delta{T}H^3}{\kappa\nu},{\quad}Pr = \frac{\nu}{\kappa},{\quad}\Gamma = \frac{L}{H}
\end{equation}
where $\alpha$, $\kappa$, and $\nu$ stand for the thermal expansion coefficient, thermal diffusivity, and kinematic viscosity of the fluid. The symbols $g$ and $\Delta{T}$ denote the gravitational acceleration and the vertical temperature difference in a fluid layer of the thickness $H$ and the characteristic horizontal dimension $L$. Convection occurs when the fluid is sufficiently heated from below and cooled from above. Warm fluid rises towards the lid and cold fluid sinks to the bottom. This leads to large-scale flows with a spatial extent in the order of the height of the container $H$. 

The flow fields show manifold patterns and can become quite complex. A large-scale circulation (LSC), which is also called the “wind of turbulence”\citep{Niemela2001, Xi2009, Krishnamurti1981, Ahlers2009}, can be observed in a wide range of parameters. There are opposing opinions in the literature whether the LSC at large $Ra$ develops directly from the steady flow patterns at small $Ra$ \citep{Busse2003, Hartlep2003} or whether this is a turbulent mode independent of it \citep{Krishnamurti1981}. The vast majority of studies to date have focused on the generic configuration of a cylinder with aspect ratio unity \citep{Ahlers2009}. Here, the LSC exists in the form of a single circulation roll. In first approximation one can assume that the single-roll LSC has a vertical planar structure where the fluid elements mainly follow an approximately circular or elliptical path \citep{Villermaux1995, Funfschilling2004, Zhou2009, Brown2009, Xi2009}. This structure usually exhibits distinct oscillations, which are attributed to torsional and sloshing modes \citep{Funfschilling2004, Zhou2009, Brown2009, Xi2009, Sun2005a, Stevens2011}. Torsion means that the transverse flow on both horizontal walls performs azimuthal oscillations, whereby a phase shift of $180^\circ$ can usually be observed between the upper and the lower half \citep{Brown2009, Zuerner2019}. This is connected with the sloshing mode which describes a gradual horizontal displacement of the circulation roll \citep{Brown2009}. The periodicity of both oscillation modes coincides with the turnover time $\tau_{to} = L_{LSC}/v_{LSC} \approx \pi H/v_{LSC}$, where $v_{LSC}$ and $L_{LSC}$ are the typical velocity magnitude and path length of the LSC, respectively \citep{Zuerner2019}. Reorientations of the LSC occur on longer time scales, which can be caused by a rotation of the LSC plane or a cessation of the flow structure \citep{Brown2005}. Azimuthal rotations of the LSC occur preferably in cylindrical geometries, because the LSC plane becomes very likely locked in a predominant orientation in the convection cell with rectangular or square cross sections \citep{Ahlers2009}. In these geometries, the LSC preferentially aligns along one or the other diagonal. Orientation changes between the two diagonal planes are also observed, which result from a lateral rotation of the LSC \citep{Bai2016}. During this, the LSC spends some finite time in a transition state in which the LSC plane is temporarily oriented parallel to a set of sidewalls \citep{Foroozani2017}.

Variations of the geometry of convection cell, in particular the aspect ratio, have a substantial impact on the large-scale structure of the flow. It turns out that the single-roll LSC represents only the special case of a large-scale flow for the cylinder with aspect ratio unity and the flow pattern becomes more complex the greater the distance from configurations with $\Gamma = 1$. Two or more rolls one above the other can exist in tall cylinders ($\Gamma < 1$) \citep{Verzicco2003, Amati2005, Sun2005b, Tsuji2005, Stringano2006, Xi2008}, while in flat containers ($\Gamma > 1$) multiple rolls occur side by side or combinations of rolls and cell structures are observed \citep{Hartlep2003, VonHardenberg2008, Yanagisawa2010, Bailon2010, Emran2015, Pandey2018, Stevens2018, Schneide2018, Sakievich2016, Sakievich2020, Krug2020}. Recent studies in shallow fluid layers at very large aspect ratios up to $\Gamma = 128$ revealed the existence of coherent flow structures that are not affected by the lateral boundary conditions. These structures survive against the background of high-frequency turbulent fluctuations for time scales being considerably longer than the turnover time during which a fluid package covers a complete circulation within the structure \citep{Emran2015, Pandey2018, Stevens2018, Schneide2018, Krug2020}. The horizontal extent of these turbulent superstructures exceeds the height of the fluid layer by several times. \citet{Pandey2018} suggested that these turbulent superstructures can be retraced up to flow patterns that are formed immediately after the onset of convection. 

Compared to the large number of studies dealing with convection in the generic aspect ratio unity or for the case of $\Gamma \gg 1$, the region of moderate aspect ratios has been sparsely studied so far. Thus, it is still an open question how the transition from a single-roll LSC to turbulent superstructures proceeds with increasing $\Gamma$. In experiments performed in air ($Pr = 0.7$), \citet{duPuits2007} observed the breakdown of the single-roll LSC to an oscillatory two-roll structure when the aspect ratio exceeds a critical value of 1.68. Further increase of $\Gamma$ beyond 3.66 causes the emergence of an unstable multi-roll regime. Both thresholds are not necessarily universal, since a dependence on $Ra$ and $Pr$ must also be taken into account. The authors suggest the second value of $\Gamma$ as a lower limit above which the turbulent convection reaches a state which becomes unaffected by the influence of the lateral walls.

Turbulent Rayleigh-Bénard convection in moderate aspect ratios has also been addressed by a few studies using direct numerical simulations (DNS). \citet{Bailon2010} investigated thermal turbulence in an air layer ($Pr = 0.7$) in a cylindrical geometry as a function of the aspect ratio in the range of $0.5 \le \Gamma \le 12$ for $Ra$ between $10^7$ and $10^9$. The authors evaluated the heat transport in connection with transitions in flow patterns. The observed large-scale flow consists of multiple rolls or cellular structures with a pentagonal or hexagonal symmetry. These patterns appear to be similar to those observed at slightly super-critical conditions close to the onset of convection for sufficiently large aspect ratios. \citet{Sakievich2016} reported the formation of hub-and-spoke structures as long-living coherent structures in a cylindrical domain at $\Gamma = 6.3$ for $Pr = 6.7$ and $Ra = 9.6\times10^7$. The flow patterns, which persist over about 600 free-fall times, appear to be similar to those obtained by \citet{Bailon2010}. The authors suggest an analogy to the "wind of turbulence" occurring in RBC at low aspect ratios. A further study by \citet{Sakievich2020} provides a continuation of the characterization of this particular regime by Fourier modal decomposition and by an extension of the total run time of the simulations covering more than 3000 free-fall times. The authors showed that compared to the standard case at $\Gamma = 1$, the flow dynamics cover much longer time scales on the order of hundreds to thousands of free-fall times. This is explained by the fact that coherent structures with larger length scales can establish themselves in increasing aspect ratio domains.

A recent experiment conducted by \citet{Vogt2018b} in a cylinder with aspect ratio $\Gamma = 2$ revealed a completely new, and until then unexpected feature of the LSC. Their measurements in liquid gallium ($Pr = 0.027$) detected a strong fluctuating flow along a measuring line that was not within the plane of the LSC. This observation cannot be reconciled with the conventional image of a quasi-two-dimensional LSC plane that only performs torsional and sloshing oscillations. Accompanying numerical simulations in $\Gamma = 2$ and $\Gamma = \sqrt{2}$ revealed a complex three-dimensional flow structure resembling a twirling jump rope. Since the existence of this jump rope vortex (JRV) has not yet been demonstrated in any other geometry, it raises the question whether this new LSC mode is a general property for RBC flows in various geometries of different aspect ratios or whether it is a special feature for a small specific range of $\Gamma.$ It is known that at aspect ratios of $\Gamma \approx 2 ... 3$ a transition range of the LSC from a single-roll regime to adjacent double rolls occurs \citep{Bailon2010}. In the vicinity of such transitions, the formation of unconventional flow structures could possibly occur.  

In this paper, we report a combined experimental and numerical work which considers the turbulent RBC in a cuboid container with square horizontal cross section of aspect ratio 5 and continues a previous experimental study made by \citet{Akashi2019}. We use the eutectic metal alloy GaInSn ($Pr$ = 0.03) as working fluid. Low $Pr$ convection is characterized by a high thermal diffusivity and an enhanced production rate of vorticity and shear which amplifies the small-scale intermittency and turbulence in the flow \citep{Scheel2017}. The dominant influence of inertia in low $Pr$ fluids qualifies them as a suitable object of investigation with respect to the formation and the dynamical behaviour of coherent flow structures in turbulent RBC. Several flow regimes were detected by \citet{Akashi2019} in the $Ra$ range  $7.9 \times 10^3 \le Ra \le 3.5 \times 10^5$. An increase of $Ra$ causes an increasing horizontal wave length of the flow structure and a conversion from a 4-roll pattern via transient 4- and 3-roll regimes to a cellular flow regime. This cellular structure is subject to pronounced regular oscillations, whose period corresponds to the turnover time, but proves to be quite stable over a long period of time. The evaluation of the temperature and velocity signals demonstrated that the flow in the investigated $Ra$ number range is turbulent. The previous study by \citet{Akashi2019} was not designed to uncover individual details of the cellular flow structure or to provide information about the changes that the structures undergo during the distinct oscillations. The motivation of the present study is to investigate the nature of these oscillations in more detail and to reveal the internal dynamics. 

The paper is organized as follows: in section 2 we present a description of the experimental setup and the numerical model. Section 3 reports on ultrasonic measurements of the flow structure and its temporal behaviour. Corresponding results obtained from numerical simulations are presented with respect to the three-dimensional flow structure, the dominating frequencies and the transport properties. A quantitative comparison of these data with the experimental findings confirm the reliability of our numerical approach. Section 4 is dedicated to the detailed analysis of the complex three-dimensional oscillations of the cellular pattern. Moreover, the consequences of the periodic changes of the flow field on the heat transport are investigated. 

\section{Experimental methods and numerical scheme}
\label{sec:exp&num}
\subsection{Experimental setup}
\label{subsec:exp-setup}

\begin{figure}
\begin{center}
   \centerline{\includegraphics[width=\linewidth]{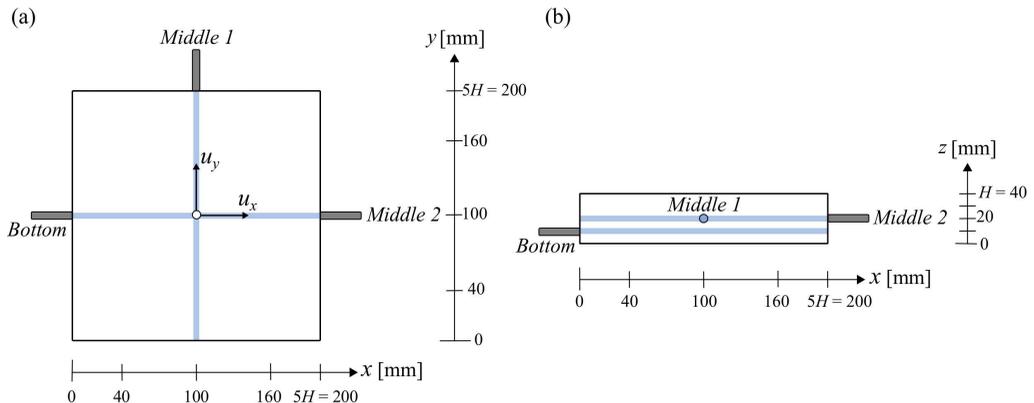}}
  \caption{Experimental setup and arrangement of measurement lines: (a) top view and (b) side view, where light blue lines indicate the ultrasonic measurement lines.}
\label{fig1}
\end{center}
\end{figure}

The setup employed for the experiments is almost identical to that used in previous studies \citep{Tasaka2016, Vogt2018a, Akashi2019, Yang2021, Vogt2021} to which the reader is referred for a detailed description. A noteworthy difference to our previous study \citep{Akashi2019} is that additional measuring positions for the velocity sensors have been placed at the mid height of the fluid container in order to better compare the flow structures found here with the results reported by \citet{Vogt2018b}. Figure \ref{fig1} shows schematics of the container and the arrangement of the ultrasonic measurement lines. The fluid layer has a square horizontal cross section of $L \times L$ = 200 $\times$ 200 mm$^2$ and a height $H$ of 40 mm, resulting in an aspect ratio of five ($\Gamma = 5$). In vertical direction the fluid layer is bordered by two copper plates, whose temperatures, $T_{bot}$ and $T_{top}$, are kept constant by water circulation in branched channels inside the plates. The water temperatures are controlled by external thermostats. The vertical temperature difference $\Delta{T}=T_{bot}-T_{top}$ is varied between 3.0 K and 15.5 K corresponding to the $Ra$ number range from $6.7 \times 10^4$ to $3.5 \times 10^5$. The side walls are made of electrically non-conducting polyvinyl chloride (PVC). The entire fluid layer is encased in 30 mm closed cell foam to minimize heat loss. The working fluid is the eutectic alloy Ga$^{67}$In$^{20.5}$Sn$^{12.5}$ that is liquid at room temperature. Temperature dependent data of the thermo-physical properties can be found in \citet{Plevachuk2014}. In particular, for a temperature of \(25~ ^\circ \)C the following values are reported for the density $\rho= 6360$ kg m$^{-3}$, the thermal expansion coefficient $\alpha = 1.24 \times 10^{-4}$ K$^{-1}$, the kinematic viscosity $\nu = 3.4 \times 10^{-7}$ m$^2$ s$^{-1}$, and the thermal diffusivity $\kappa = 1.03 \times 10^{-5}$ m$^2$ s$^{-1}$. This gives a $Pr$ number of $Pr = 0.033$. The thermal diffusion time for the fluid layer, $t_\kappa = H^2/\kappa$, is $155$ s. Table \ref{tab1} presents the $Ra$ investigated in the experiments together with the corresponding values for the free-fall velocity $u_{ff}= \sqrt{g\alpha\Delta{T}H}$ and the free-fall time $t_{ff} =\sqrt{H/(g\alpha\Delta{T})}$ resulting from the respective temperature difference $\Delta{T}$.

Due to the limited possibilities for measuring fluid velocities in liquid metals, the majority of previous experimental studies on liquid metal convection were largely restricted to the description of the heat transfer in the system and the measurement of temperatures and their fluctuations \citep{Takeshita1996, Cioni1997, Segawa1998, Horanyi1999, Burr2001}. The development of the ultrasonic Doppler technique for flow measurements in liquid metals \citep{Takeda1987, Eckert2002} enables an experimental determination of flow structures in low $Pr$ liquid metal convection \citep{Cramer2004, Mashiko2004, Yanagisawa2010, Yanagisawa2013,Tasaka2016,  Zuerner2019, Tasaka2021, Yang2021, Vogt2021}. This powerful technique was successfully used in the previous study by \cite{Akashi2019} to identify the individual flow regimes and is also employed here. It provides instantaneous profiles of the velocity component projected onto each measurement line, $u_x(x,t)$ and $u_y(y,t)$, respectively. The spatial resolution of the velocity measurements is 1.4 mm in the direction of the ultrasonic beam line and about 5 to 8 mm in the direction perpendicular to it. The velocity resolution is about 0.5 mm s$^{-1}$ and a temporal resolution of 0.6 s is achieved. 
The measurement instrument DOP 3010 (Signal Processing S.A.) is applied in combination with ultrasonic transducers having a basic frequency of 8 MHz and a piezo element with an active diameter of 5 mm. These sensors, which are mounted horizontally inside holes drilled into the sidewalls of the container, are in direct contact with the liquid metal. In general, velocity probes can be installed at side walls at different heights. In the present study, measurement results are obtained along three measurement lines (shown as light blue lines in figure \ref{fig1}), designated as $Middle \textit{1}$, $Middle \textit{2}$ and $Bottom$. $Middle \textit{1}$ and $Middle \textit{2}$ are at the mid height of the container ($z = 0.5H$), and $Bottom$ is situated close to the bottom plate ($z = 0.25H$). The velocity profiles along each measurement line are acquired sequentially by multiplexing.

\subsection{Numerical scheme}
\label{subsec:num-model}

For the numerical simulation, a Boussinesq fluid in a cuboid container with a square horizontal cross section and an aspect ratio 5 is considered. Here, we used the nonmagnetic version of a code originally developed for magnetohydrodynamic (MHD) simulations in \citet{Yanagisawa2015}. Cartesian coordinates $(x, y, z)$ are used with the $z$-axis in the upward direction and the origin located at one of the bottom corners of the container. We solve the governing equations for the non-dimensional velocity \mbox{\boldmath$u$}, the temperature $T$, and pressure $p$ as follows: 
\begin{equation}
\frac{\partial \mbox{\boldmath$u$}}{\partial t}=-(\mbox{\boldmath$u$}\cdot\nabla)\mbox{\boldmath$u$}-\nabla p+PrRa (\mbox{\it{T}}-\overline{T}(z)) \mbox{\boldmath$k$}_z+Pr\nabla^2\mbox{\boldmath$u$}  , \:  \mbox{ }\overline{T}(z)=1-z ,
\end{equation}
\begin{equation}
\nabla\cdot\mbox{\boldmath$u$}=0 ,
\end{equation}
\begin{equation}
\frac{\partial T}{\partial t}=-(\mbox{\boldmath$u$}\cdot\nabla)T+\nabla^2 T , 
\end{equation}
where $\mbox{\boldmath$k$}_z$ is the unit vector in the $z$ direction. Length and time are respectively nondimensionalized by the layer thickness, $H$, and the thermal diffusion time, $t_\kappa=H^2/\kappa$. Accordingly, the velocity is normalized by $\kappa/H$. The temperature satisfies the isothermal boundary conditions at the top ($T_{top}$ = 0) and bottom ($T_{bot}$ = 1), while adiabatic conditions are assumed for the sidewalls. No-slip conditions are applied for the velocity at all boundaries. The second-order accurate staggered-grid finite difference method is used with a uniform grid interval in each direction. The time integration was carried out explicitly by means of the third-order Runge-Kutta method. This procedure modifies the equation of continuity to 
\begin{equation}
\varepsilon^2{\partial p}/{\partial t}=-\nabla\cdot\mbox{\boldmath$u$}, 
\end{equation} 
where $\varepsilon$ is a small parameter from \citet{Chorin1967}. In our numerical simulations a value of $\varepsilon=0.002$ is used \citep{Yanagisawa2015}. The code is parallelized with MPI in combination with OpenMP.

To check the resolution of the simulation, we compared results from calculations made with different grid resolutions $n_z = 80, 128, 256$, and $512$) (see table \ref{tab2}). For these pre-calculations, the parameters are set as $Pr = 0.025$, which is a representative value for liquid metals, and $Ra = 1.0\times 10^5$, $Ra = 3.0\times 10^5$ and $Ra = 6.0\times 10^5$. All simulations performed were able to reconstruct the same qualitative characteristics of the flow field: a cellular pattern with quasi-periodic oscillations. Quantitative comparisons are carried out using the values of the Nusselt number and the Reynolds number which were averaged both temporally and spatially. In our equations $\langle \cdot \rangle_{S, t}$ stands for the time--surface average on the top or bottom boundary, and $\langle \cdot \rangle_{V, t}$ stands for the time--volume average. Nusselt numbers are calculated both at the top boundary $Nu_{top}$ and at the bottom boundary $Nu_{bot}$
\begin{equation}
Nu_{top}=-\left \langle \left (\frac{\partial T}{\partial z} \right)_{z=1} \right \rangle_{S, t}, \:\:\:Nu_{bot}=-\left \langle \left(\frac{\partial T}{\partial z} \right)_{z=0} \right \rangle_{S, t}.
\label{Nu1}
\end{equation}
An alternative way to determine the Nusselt number is to integrate the upward heat transport in the entire volume (e.g. \citet{Pandey2018}). For comparison we also use this definition for the Nusselt number $Nu_{vol}$

\begin{equation} 
Nu_{vol}=1\: +\langle u_z T \rangle_{V, t}.  
\label{Nu2}
\end{equation}

When thermally balanced states are achieved after adequate time integrations, applying equations (\ref{Nu1}) and (\ref{Nu2}) to our data yields a very good agreement. Therefore, in the further course of the paper only the values of $Nu_{top}$ will be presented as $Nu$.

The Reynolds number is calculated using the root mean square (rms) velocities $U_{rms}$ determined for the entire volume as characteristic velocity scale

\begin{equation}
 Re = \frac{U_{rms}}{Pr} \:\:\: \mbox{with} \:\:\: U_{rms}=\:\sqrt{\langle {u_x}^2+{u_y}^2+{u_z}^2 \rangle_{V, t}}. 
\label{Re_num}
\end{equation}

\begin{table}
\begin{center}
\def~{\hphantom{0}}
\color{black}
\begin{tabular}{ccccccccccc}
       ~$Ra$~        &   $Pr$    &   $u_{ff}[mm/s]$   &    $t_{ff}[s]$  \\[3pt]
   ~$6.7\times10^4$~ &  ~0.033~  &    ~12.28~         &    ~3.28~       \\
	 ~$8.6\times10^4$~ &  ~0.033~  &    ~13.84~         &    ~2.88~       \\
	 ~$9.0\times10^4$~ &  ~0.033~  &    ~14.19~         &    ~2.82~       \\
	 ~$1.0\times10^5$~ &  ~0.033~  &    ~15.43~         &    ~2.59~       \\
	 ~$1.2\times10^5$~ &  ~0.033~  &    ~16.48~         &    ~2.43~       \\
	 ~$1.5\times10^5$~ &  ~0.033~  &    ~18.40~         &    ~2.17~       \\
   ~$1.8\times10^5$~ &  ~0.033~  &    ~19.67~         &    ~2.03~       \\
	 ~$2.2\times10^5$~ &  ~0.033~  &    ~21.98~         &    ~1.81~       \\
	 ~$2.6\times10^5$~ &  ~0.033~  &    ~24.00~         &    ~1.67~       \\
	 ~$3.5\times10^5$~ &  ~0.033~  &    ~27.84~         &    ~1.43~       \\
  \end{tabular}
  \caption{Parameters for the experiments including the Rayleigh number $Ra$, the Prandtl number $Pr$, the free-fall velocity $u_{ff}$, and the free-fall time $t_{ff}$.}
  \label{tab1}
  \end{center}
\end{table}

\begin{table}
\begin{center}

\def~{\hphantom{0}}
\begin{tabular}{ccccccccccc}
       ~$Ra$~        &   $Pr$    &  $n_x$   &  $n_y$  &  $n_z$  &  $Nu$   & $Nu_{std}$& $n_{\lambda_{\Theta}}$&   $Re$   & $Re_{std}$ & $\eta_{K}/l_z$ \\[3pt]
   ~$1.0\times10^5$~ &  ~0.025~  &  ~400~   &  ~400~  &  ~80~   & ~2.76~  &  ~0.08~   &       ~14~            &  ~1011~  &   ~18.3~   &     ~0.62~     \\
	 ~$1.0\times10^5$~ &  ~0.025~  &  ~640~   &  ~640~  &  ~128~  & ~2.76~  &  ~0.09~   &       ~23~            &  ~1009~  &   ~22.4~   &     ~0.99~     \\
	 ~$1.0\times10^5$~ &  ~0.025~  &  ~1280~  &  ~1280~ &  ~256~  & ~2.77~  &  ~0.07~   &       ~46~            &  ~1011~  &   ~16.2~   &     ~1.97~     \\
	 ~$1.0\times10^5$~ &  ~0.025~  &  ~2560~  &  ~2560~ &  ~512~  & ~2.76~  &  ~0.07~   &       ~93~            &  ~1009~  &   ~16.0~   &     ~3.95~     \\
	 
	 ~$3.0\times10^5$~ &  ~0.025~  &  ~400~   &  ~400~  &  ~80~   & ~3.73~  &  ~0.13~   &       ~10~            &  ~1706~  &   ~27.2~   &     ~0.42~     \\
	 ~$3.0\times10^5$~ &  ~0.025~  &  ~640~   &  ~640~  &  ~128~  & ~3.74~  &  ~0.12~   &       ~17~            &  ~1708~  &   ~26.8~   &     ~0.67~     \\
	 ~$3.0\times10^5$~ &  ~0.025~  &  ~1280~  &  ~1280~ &  ~256~  & ~3.73~  &  ~0.14~   &       ~34~            &  ~1708~  &   ~25.7~   &     ~1.35~     \\
	 ~$3.0\times10^5$~ &  ~0.025~  &  ~2560~  &  ~2560~ &  ~512~  & ~3.73~  &  ~0.14~   &       ~69~            &  ~1706~  &   ~26.5~   &     ~2.70~     \\
   
	 ~$6.0\times10^5$~ &  ~0.025~  &  ~400~   &  ~400~  &  ~80~   & ~4.50~  &  ~0.16~   &       ~9~             &  ~2317~  &   ~36.2~   &     ~0.33~     \\
	 ~$6.0\times10^5$~ &  ~0.025~  &  ~640~   &  ~640~  &  ~128~  & ~4.51~  &  ~0.14~   &       ~14~            &  ~2335~  &   ~34.2~   &     ~0.53~     \\
	 ~$6.0\times10^5$~ &  ~0.025~  &  ~1280~  &  ~1280~ &  ~256~  & ~4.51~  &  ~0.14~   &       ~28~            &  ~2334~  &   ~37.5~   &     ~1.06~     \\
	 ~$6.0\times10^5$~ &  ~0.025~  &  ~2560~  &  ~2560~ &  ~512~  & ~4.51~  &  ~0.16~   &       ~57~            &  ~2234~  &   ~39.9~   &     ~2.13~     \\
  \end{tabular}
  \caption{Pre-tests of numerical simulations conducted with different numbers of grid points: the Rayleigh number $Ra$, the Prandtl number $Pr$, the number of grid points $n_x$, $n_y$, and $n_z$, the time-averaged Nusselt number $Nu$, the standard deviation of Nusselt number $Nu_{std}$, the number of grid points in the thermal boundary layer $\lambda_{\Theta}$ estimated by the relation, $Nu = 1/{(2\lambda_{\Theta})}$, $n_{\lambda_{\Theta}}$, the time-averaged Reynolds number $Re$, the standard deviation of Reynolds number $Re_{std}$, and the ratio of the Kolmogorov scale $\eta_{K}$ to the spatial resolution $l_z$.}
  \label{tab2}
  \end{center}
\end{table}

\begin{table}
\begin{center}

\def~{\hphantom{0}}
\begin{tabular}{ccccccccccc}
       ~$Ra$~        &   $Pr$    &   $n_x$   &  $n_y$  &  $n_z$  &  $u_{ff}[mm/s]$ &  $t_{ff}[s]$ \\[3pt]
   ~$1.0\times10^5$~ &   0.03    &   ~640~   &  ~640~  &  ~128~  &  ~13.13~        &  ~3.05~      \\
	 ~$1.2\times10^5$~ &   0.03    &   ~640~   &  ~640~  &  ~128~  &  ~15.75~        &  ~2.54~      \\
	 ~$1.2\times10^5$~ &   0.03    &   ~400~   &  ~400~  &  ~80~   &  ~15.75~        &  ~2.54~      \\
	 ~$2.0\times10^5$~ &   0.03    &   ~640~   &  ~640~  &  ~128~  &  ~20.33~        &  ~1.97~      \\
	 ~$3.0\times10^5$~ &   0.03    &   ~640~   &  ~640~  &  ~128~  &  ~24.90~        &  ~1.61~      \\
	 ~$6.0\times10^5$~ &   0.03    &   ~640~   &  ~640~  &  ~128~  &  ~35.22~        &  ~1.14~      \\
  \end{tabular}
  \caption{Parameters for the numerical simulations of the experimental configuration including the Rayleigh number $Ra$, the number of grid points $n_x$, $n_y$, and $n_z$, the free-fall velocity $u_{ff}$, and the free-fall time $t_{ff}$.}
  \label{tab3}
  \end{center}
\end{table}

The quantitative comparison of the pre-tests is shown in table \ref{tab2}. The agreement of the time-averaged Nusselt numbers $Nu$ is satisfactory within 0.4\%, and the Reynolds numbers $Re$ are almost the same within 0.8\%. As will be shown below, quasi-periodicity is a dominant feature of the pattern, hence, fluctuations of the Nusselt number are also important. The fluctuations are evaluated in form of the standard deviation and are listed in table \ref{tab2} as $Nu_{std}$. 

In addition, we checked the spatial resolution of our simulations by comparing them with the Kolmogorov scale $\eta_{K}$. Using the results of $Nu$ for given values of $Pr$ and $Ra$, $\eta_{K}$ can be estimated as \citet{Scheel2013}.
\begin{equation}
\eta_{K} =\left(\frac{Pr^2}{(Nu-1)Ra}\right)^{\frac{1}{4}}. 
\label{kolm}
\end{equation}
Table \ref{tab2} contains the ratio of $\eta_{K}$ to the spatial resolution $l_z= 1/n_z$. A ratio larger than one indicates a sufficient resolution in terms of Direct Numerical Simulation (DNS), hence, these values are used as references. As shown in table 2 simulations with the highest grid resolution satisfy this criterion for the $Ra$ range considered here. Calculations of lower resolution are conducted in the case of long simulation times. Due to the high diffusivity of temperature, we assume that this is nevertheless sufficient to adequately evaluate $Nu$, $Re$, temperature distribution and flow pattern. Our independence grid study shows that the values of the quantities we focus on in this paper do not change significantly on a finer grid (see above), so we can assume their validity. Higher resolutions of the grid are necessary if details within the velocity boundary layers are to be examined, but this is outside the scope of this study.

Based on these comparisons and considering available computational resources, we performed simulations mainly with the resolution $n_z=128$. To check the long term stability of the quasi-periodic behaviour and to perform longtime averaging over 100 thermal diffusion times ($\sim 300$ oscillations), we utilized the resolution of $n_z=80$. The parameters chosen for the numerical simulations in this study are listed in table \ref{tab3}. The numerical simulations cover the parameter range that is investigated in the experiments.

\section{Characterization of the large-scale flow structure}
\label{sec:3D-temp}

\subsection{Three-dimensional cellular regime}
\label{subsec:flow-patte}

Previous experiments carried out by \citet{Akashi2019} in the same setup, show a clear and reproducible dependence on $Ra$ with respect to the occurrence of different flow regimes. While the range of smaller $Ra$ is associated with various regimes of unsteady convection rolls, at $Ra$ above $6.5\times10^4$ a new, stable flow pattern occurs, which is characterized by a high symmetry with respect to the square base area of the fluid container. This newly discovered flow pattern was designated as the cellular flow regime. We are aware that the term of a cellular flow structure in thermal convection is often associated with stationary cell patterns observed, for example, in the slightly supercritical state after the onset of convection. This is explicitly not the case here. Accompanying measurements of the temperature fluctuations indicate that the stage of developed thermal turbulence is reached in the $Ra$ range, where the cellular regime occurs \citep{Akashi2019}. Our point of view is supported by results from  \citet{Bailon2010} who found extended convection rolls and pentagon-like cells in their DNS data for a turbulent flow in a $\Gamma = 12$ cylinder at $Ra = 10^7$ and $Pr = 0.7$. After filtering out the small-scale turbulence, these resulting patterns of turbulent convection resemble the weakly nonlinear regime just above the onset of convection. However, the observation of the cellular regime by \citet{Akashi2019} is also astonishing because the appearance of such cell-like structures is anticipated to be more prevalent in fluids with larger $Pr \gtrsim 1$, while the convection pattern at low $Pr$ is supposed to be dominated by roll-like structures \citep{Breuer2004, Pandey2018}. In order to analyze the cellular flow regime in more detail, the investigations in this study are focused on the $Ra$ number range $6.7 \times 10^4 \leq Ra \leq 3.5 \times 10^5$. The upper limit of $Ra$ is determined by the technical capabilities of the experimental equipment.   

\begin{figure}
\begin{center}
  \centerline{\includegraphics[width=\linewidth]{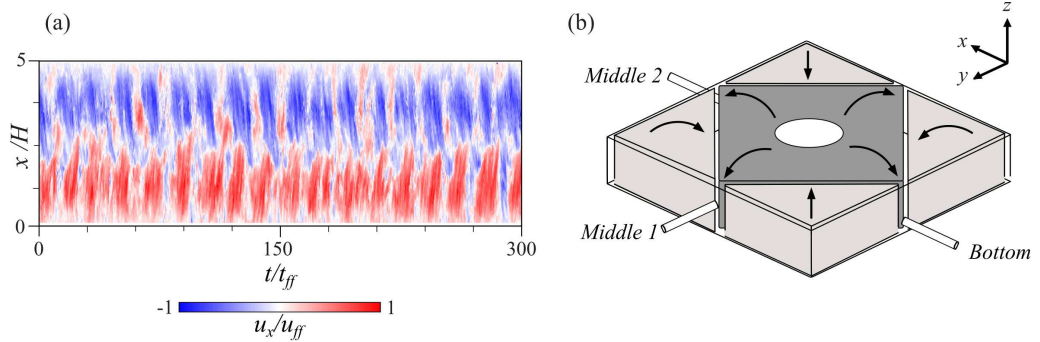}}
  \caption{(a) Spatio-temporal velocity map at $Ra=1.2\times10^5$ recorded for $u_{x}$ along the $x$ direction (sensor $Bottom$ at $z = 0.25H$), (b) a schematic illustration of the cell structure with the sensor positions.}
\label{fig2}
\end{center}
\end{figure}

Figure 2(a) shows the spatio-temporal map of the velocity component $u_x$ at $Ra=1.2\times10^5$ recorded by the ultrasonic sensor $Bottom$ at a height of $z = 0.25 H$. A negative velocity (blue) indicates a flow toward the transducers, while the positive velocity (red) is associated with the direction away from the sensor. In all results presented in the following, the velocity is given dimensionless with respect to the free-fall velocity $u_{ff}= \sqrt{g\alpha\Delta{T}H}$. In the convection cell considered here, the free-fall velocity reaches a value of 16.48 mm/s at $Ra=1.2\times10^5$. The dimensionless time is given in units of the free-fall time $t_{ff} =\sqrt{ H/(g\alpha\Delta{T})} = 2.43$ s.

The velocity map obtained by the sensor $Bottom$ (figure 2(a)) reproduces an almost identical flow structure as those presented in the recent study by \citet{Akashi2019}. The flow pattern divides the measured profile into two parts with respect to the centre of the container. The strength of the measured velocity is subject to strong quasi-periodic fluctuations, which at some points in time (e.g. $t/t_{ff} = 100, 150, 270$) even lead to short-term reversals of the flow direction. If the average behaviour of the flow is considered, then a fluid motion from the side wall towards the centre is clearly dominant, which means that an ascending flow in the centre of the cell can be assumed. Based on the evaluation of the measured velocity profiles along 6 different measurement lines within a horizontal plane above the heated bottom, \citet{Akashi2019} obtained a rough reconstruction of the time-averaged velocity patterns in the convection cell. For the case of the cellular regime, they revealed a three-dimensional structure having upwelling flows at the centre and the four corners of the container as shown in figure \ref{fig2}(b) \citep{Akashi2019}. Once this flow pattern is established, it is stable for long measurement times in the order of a few thousand free fall times and does not change spontaneously. Repeating the measurements several times, inverse flow patterns with a downward flow in the centre of the fluid container were also observed. These structures, which differ principally in the sign of the velocity, have similar probabilities of occurrence. Which variant is realized in the experiment apparently depends only on random asymmetries in the initial conditions.    

\begin{figure}
\begin{center}
  \centerline{\includegraphics[width=\linewidth]{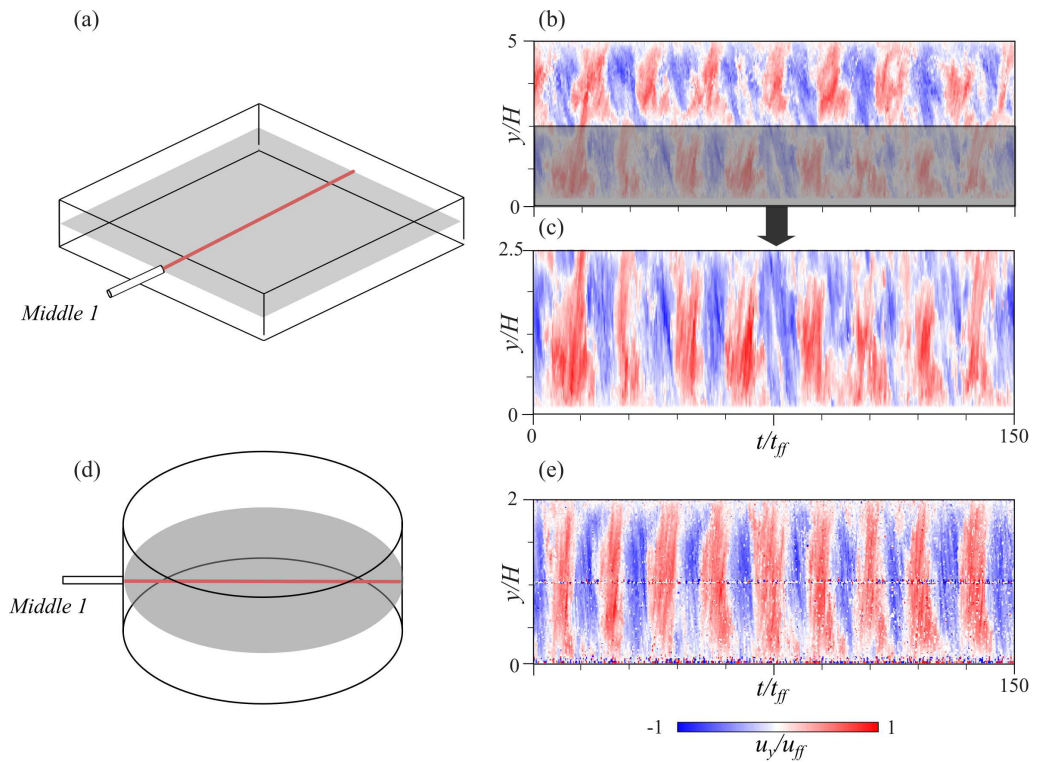}}
  \caption{Spatio-temporal velocity distributions measured at the mid-height of the fluid container for $Ra = 1.2 \times 10^5$: (a) sketch of the measuring line in the $\Gamma$ = 5 box ($Pr$ = 0.033), (b) velocity pattern for the the full length measuring line (c) the same data set as in (b) plotted for only one half of the measuring line, (d) sketch of the measuring line in the $\Gamma$ = 2 cylinder ($Pr$ = 0.027), (e) velocity pattern in the $\Gamma$ = 2 cylinder (adopted from \citet{Vogt2018b}).}
\label{fig2plus}
\end{center}
\end{figure}

Inspired by the work of \citet{Vogt2018b}, here we additionally monitor the velocities in the horizontal centre plane of our container. If the convection pattern corresponds to a single circulation, which fills the space between the copper plates, one would expect that in the measuring plane at the mid-height the vertical component clearly dominates and only marginal parts of a horizontal flow can be found here. As shown in figure \ref{fig2plus}(b) and (c), it is surprising to observe pronounced flows also in the centre plane, their intensity is only slightly below that detected along the measuring lines at $z = 0.25H$ and $z = 0.75H$. Moreover, these velocity plots are characterized by the same pronounced oscillations. In contrast to the measurement at $z = 0.25H$, however, an inversion of the flow direction occurs in every period, where no dominance of one of the two directions can be determined. Figure \ref{fig2plus} also contains a corresponding data set from \citet{Vogt2018b} (see figure \ref{fig2plus}(d) and (e)). For better comparability, the time axis and the velocity are normalized with the free-fall time and the free-fall velocity, respectively. The comparison of both flow patterns shows striking similarities with respect to the manifestation of the dominant oscillations both in terms of the amplitude of the velocity and the time scale of the oscillations. This raises the question whether we are dealing with the same phenomenon in the $\Gamma=5$ box that is referred to as JRV and has already been observed in the $\Gamma=2$ cylinder. To answer this question, we take a closer look at the flow structure and its changes during the oscillations. Since the experiment reaches its limits here due to the small numbers of measuring sensors, we performed numerical simulations in parallel. 

\begin{figure}
\begin{center}
  \centerline{\includegraphics[width=\linewidth]{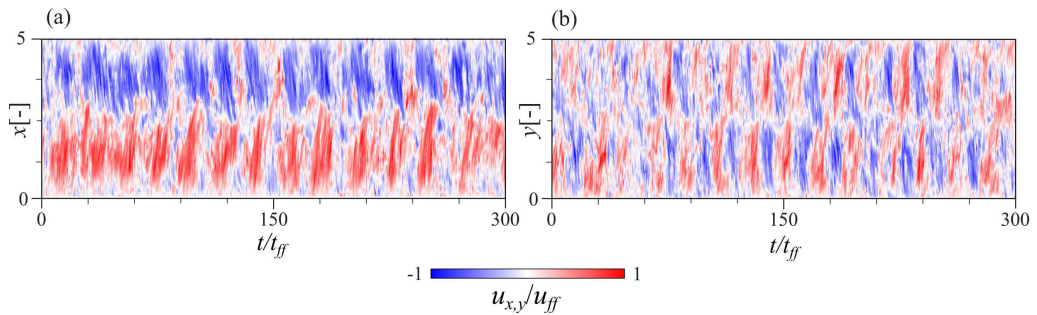}}
  \caption{Results of the numerical simulations showing a reconstruction of the spatio-temporal velocity maps at $Ra=1.2\times10^5$ in the cuboid container for (a) $u_x$ according to the measuring line of sensor $Bottom$ at $z = 0.25$, (b) $u_y$ according to the measuring line of sensor $Middle \textit{1}$ at $z = 0.5$. The measurement positions are shown in figure \ref{fig2}(b).}
\label{fig3}
\end{center}
\end{figure}

Figure \ref{fig3}(a) and (b) contain the numerical counterpart of the measurements as presented in figure \ref{fig2} and figure \ref{fig2plus}. The qualitative agreement of the flow patterns proves the reliability of the numerical approach. The numerical velocity distributions are also dominated by strong oscillations, with periodic changes of direction in the centre plane of the container.

\begin{figure}
\begin{center}
		\centerline{\includegraphics[width=\linewidth]{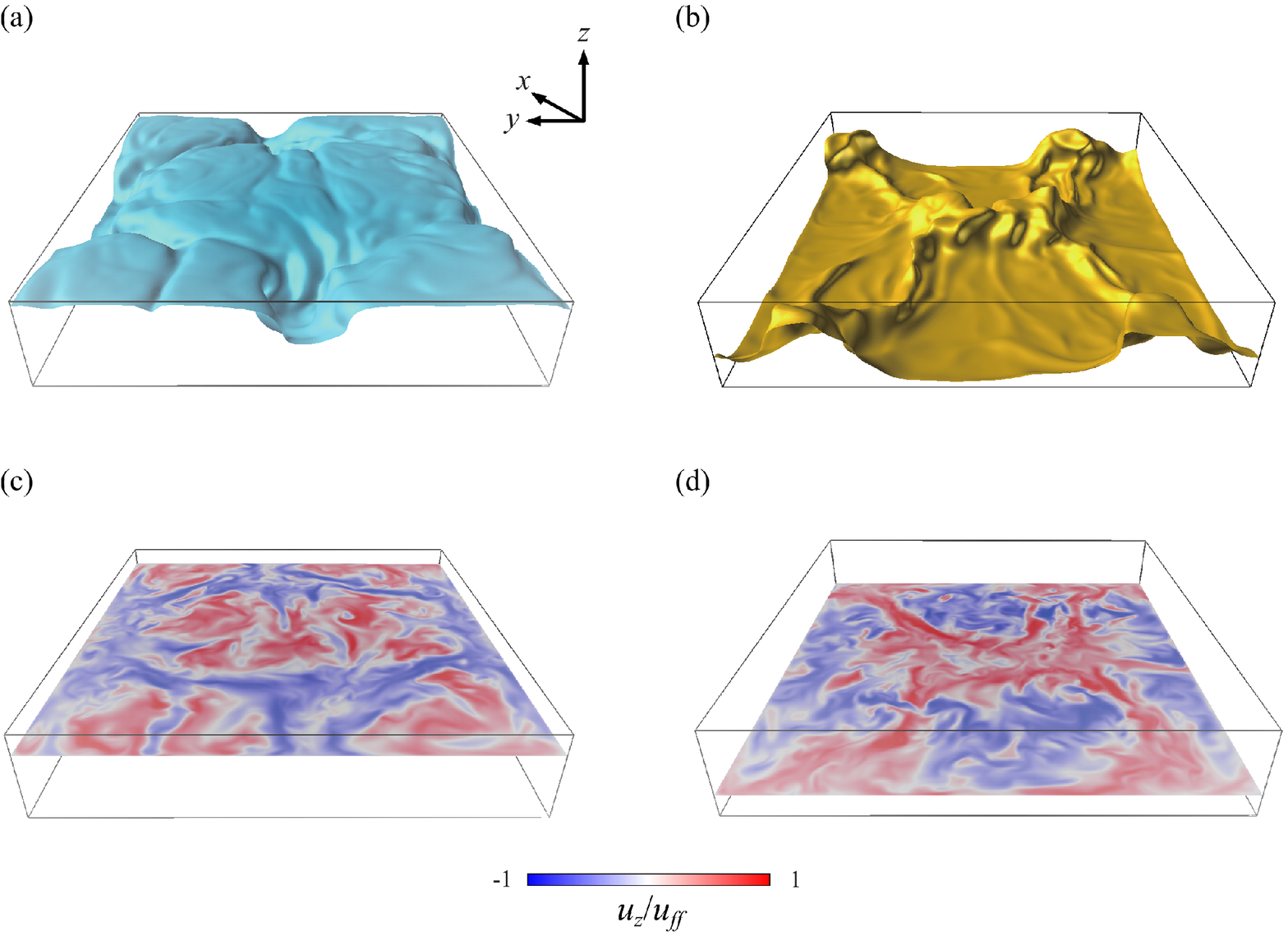}}
		\caption{Results of the numerical simulations performed for $Ra=1.2\times10^5$ representing snapshots of the temperature isosurface for (a) $T=0.2$, (b) $T=0.8$, and the distribution of the vertical velocity $u_z$ at the height of (c) $z = 0.75$, (d) $z = 0.25$. The colour scale of $u_z$ is same as that in figure 4. See also supplementary materials Movie 1-4.}
\label{fig4}
\end{center}
\end{figure}

Snapshots of temperature isosurfaces and velocity distributions in figure \ref{fig4} provide a three-dimensional visualization of the convection pattern. Figures \ref{fig4}(a) and (b) display instantaneous isosurfaces of the temperature for the values $T = 0.2$ and $T = 0.8$. Figures \ref{fig4}(c) and (d) show the instantaneous distribution of the vertical velocity $u_z$ at different heights of $z = 0.75$ and $z = 0.25$. The colour indicates the direction of flow, that means that blue (red) signifies a downwelling (upwelling) motion. The temperature surface appears to be smoother than the velocity fields, which is naturally due to the fact that the low $Pr$ number fluids feature a high thermal diffusivity. Apart from that, temperature field and vertical velocity show an evident conformity. Previous studies on turbulent superstructures raised the question to what extent the temperature distribution and the vertical velocity structure represent the same coherent structure \citep{Pandey2018, Stevens2018, Krug2020}. \citet{Sakievich2016} showed that the thermal updrafts and downdrafts match the three-dimensional field to a high degree in a $\Gamma=6.3$ cylinder. Further analysis by \citet{Sakievich2020} revealed a high correlation in the temporal dynamics of the low-order Fourier modes of temperature and vertical velocity. Based on this clear similarity and the findings by \citet{Krug2020}, \citet{Sakievich2020} hypothesize similar principles of spatial organization of the coherent patterns in the $\Gamma=6.3$ cylinder and superstructures occurring in larger aspect ratios.

The temperature plots in figure \ref{fig4} show that the cellular structure is characterized by an ascending flow in its centre and at the four corners of the container (see figure \ref{fig4}(b)), while the shape of the areas where the fluid descends resembles a rhombus (see figure \ref{fig4}(a)). It can also be seen that the zones of ascending warm fluids in the middle and the four corners are connected by diagonally running ridges (figure \ref{fig4}(b) and \ref{fig4}(d)). This rough characterization of the flow structure is consistent with the distributions of the vertical velocity component in the figures \ref{fig4}(c) and \ref{fig4}(d). The appearance of finer structures in the velocity contours illustrates the character of inertia-dominated fluid turbulence and is in line with previous studies \citep{Pandey2018, Vogt2018b, Krug2020, Sakievich2020} which demonstrated the occurrence of more small-scale velocity fluctuations compared to the temperature field. Thus, it can be stated that the results of the numerical simulations confirm the mean flow structure of the cellular regime suggested from the experiment based on a limited number of velocity sensors \citep{Akashi2019}. A striking property of this cellular convection is the occurrence of pronounced periodic oscillations, which consequently must be reflected in some form of periodic changes of the flow topology. 

\subsection{Dominant oscillation frequencies}
\label{subsec:Osci-freq}

\begin{figure}
\begin{center}
  \centerline{\includegraphics[width=\linewidth]{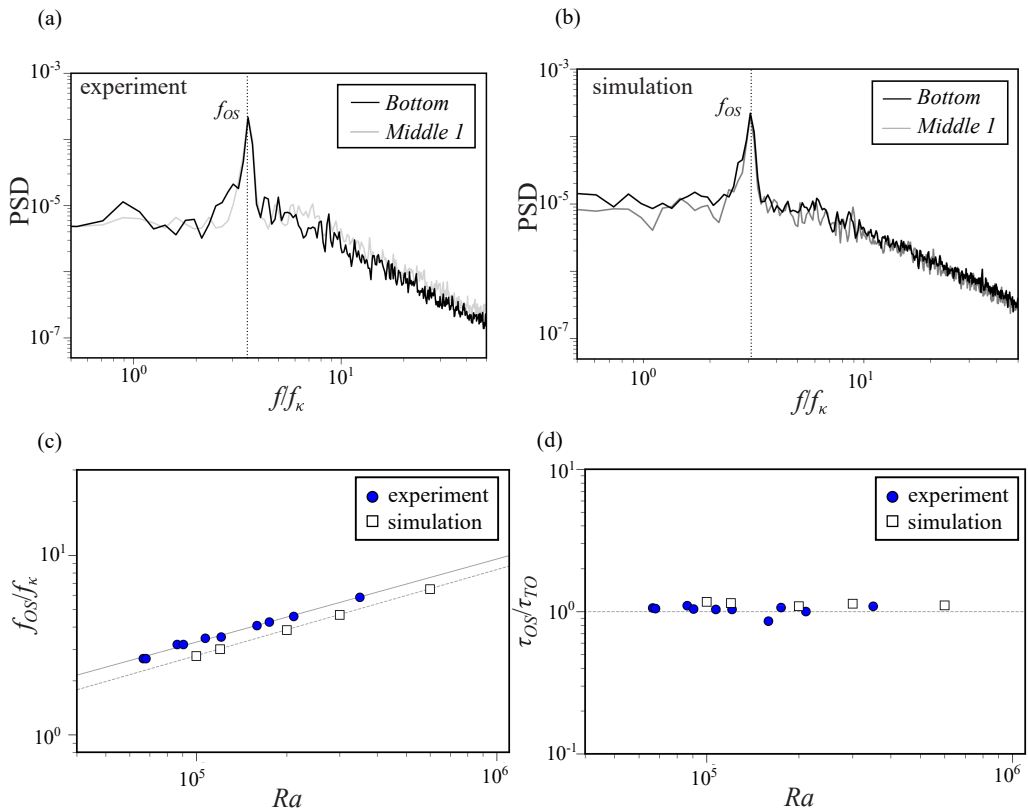}}
  \caption{Spatially averaged PSDs calculated from the spatio-temporal velocity maps for $Ra = 1.2\times10^5$ (a) measured in the experiments by the sensors $Bottom$ and $Middle \textit{1}$,(b) obtained from numerical simulations, (c) oscillation frequency $f_{OS}$ as a function of $Ra$ from experimental and numerical data, where the solid line represents the least-square fit of the experimental results, $f_{OS}H^2/\kappa= 0.016Ra^{0.46}$ and the dotted line shows the fitting of the results of numerical simulations, $f_{OS}H^2/\kappa= 0.011Ra^{0.48}$, and (d) the period length of the oscillations normalized with the turnover time $\tau_{TO}$ versus $Ra$, where the the dotted line represents a value of one.}
\label{fig5}
\end{center}
\end{figure}

The dominant frequency $f_{OS}$ of the periodic oscillations in the spatio-temporal velocity maps is determined by calculating the spatially averaged power spectral densities (PSDs) from velocity time series. The velocity time series are divided into multiple time intervals of 60 free-fall times. The spectra are calculated by Fast Fourier transform (FFT) for each time interval and for each measurement point. These interim results are finally averaged both spatially along the respective measuring line and temporally over the duration of the measurement. 

Figures \ref{fig5}(a) and (b) present the PSDs obtained from the velocity maps along the measurement lines covered by the sensors $Bottom$ and $Middle \textit{1}$. The frequency is normalized by the thermal diffusion frequency $f_\kappa = \kappa/H^2$.
The dominant oscillation frequency can be easily identified by the clear peaks in the PSDs. The same frequency is found at both measuring positions indicating a certain coherence of the flow structure. The spectra in both figures \ref{fig5}(a) and (b) show similar magnitudes over the whole frequency range and an identical slope in the inertial domain for $f/f_\kappa > 10^1$. There is a slight deviation between experiment and numerical simulation concerning the values of the normalized frequency $f/f_\kappa = 3.3$ and $f/f_\kappa = 3.1$, respectively. 

The oscillation frequency $f_{OS}$ increases with increasing $Ra$ as shown in figure \ref{fig5}(c). A power-law fit leads to a scaling of $f_{OS}/f_\kappa =0.016Ra^{0.46\pm0.02}$ for the velocity measurements in the experiments and $f_{OS}/f_\kappa =0.011Ra^{0.48\pm0.02}$ for the numerical data. The agreement is satisfactory, where the exponents agree within the error bars. The slight discrepancy may result from uncertainties in the material properties, which in turn could affect the exact determination of the $Pr$ and $Ra$ in the experiment. In the paper by \citet{Plevachuk2014}, from which the numerical values for the thermal diffusivity, viscosity and density of GaInSn were taken, the accuracy of these measured values is specified in the range from $\pm1.5\%$ to $\pm7\%$, which might be sufficient to explain the above mentioned deviations.   

Previous results reported by \citet{Akashi2019} for experiments in the same container reported a scaling law of $f_{OS}/f_\kappa= 0.031Ra^{0.40\pm 0.02}$. The discrepancy can be explained by the fact that the range in which the power-law fit was calculated in \citet{Akashi2019} also includes smaller $Ra$, where unstable rolls occur instead of the cellular structure. This also indicates that the two separate flow patterns cause different exponents in the scaling law. Experimental \citep{Tsuji2005, Vogt2018b, Zuerner2019} and numerical studies \citep{Scheel2017} in cylindrical cells with $\Gamma = 1$ and $\Gamma =2$ come to smaller exponents. \citet{Zuerner2019} found $f_{OS}/f_\kappa= (0.010\pm)\times Ra^{0.40\pm 0.02}$ at $Pr = 0.029$. The DNS by \citet{Schumacher2016} at $Pr=0.021$ results in a scaling $f_{OS}/f_\kappa= (0.08\pm)\times Ra^{0.42\pm 0.02}$. Another scaling law, $f_{OS}/f_\kappa= (0.027\pm)\times Ra^{0.419\pm 0.006}$, is given by \citet{Vogt2018b}. 

\citet{Akashi2019} demonstrated that the ratio between the oscillation period $\tau_{OS} = 1/f_{OS}$ and the turn over time $\tau_{TO}$ is close to unity. The turn over time $\tau_{TO}$ describes the time required for a fluid parcel to complete a full circulation within the LSC structure. Except for the movement in the corners, the fluid elements follow in good approximation the path of an ellipse with the semi-axes $L/4$ and $H/2$. Using a simple approximation formula to determine the circumference of an ellipse, the turnover time can be calculated as follows
\begin{equation}
\tau_{TO} = \frac{\pi(\sqrt{2((L/4)^2 + (H/2)^2)}}{u_{rms}}.
\end{equation}
The velocity $u_{rms}$ is calculated from the velocity data recorded by the sensor $Bottom$ as
\begin{equation}
u_{rms}=\overline{\sqrt{\frac{1}{L}\int^{L}_0u_x^2(x,t)dx}}. 
\end{equation}
The ratio $\tau_{OS}$ to $\tau_{TO}$ is plotted versus $Ra$ in figure \ref{fig5}(d). All data points are in a close proximity to the value one. 

The occurrence of characteristic frequency peaks is observed in many experimental studies \citep{Qiu2001,  Castaing1989, Siggia1994} and is apparently a characteristic feature of coherent large-scale structures occurring in thermal convection. Several studies construe these oscillations as an indication of the transport of individual plumes by the LSC \citep{Cioni1997, Grossmann2000, Grossmann2002, Stevens2013}. Another perspective is to associate these oscillations with the advection of deviations from an ideal two-dimensional LSC structure or other substructures by the LSC. This in a natural way explains an inherent relationship between the frequency of the oscillations and the turnover time. Thus, the torsion and sloshing phenomenon could also be understood as a deformation of the two-dimensional single-roll LSC that travels with the recurrence rate $1/\tau_{TO}$ throughout the fluid container \citep{Zuerner2019}. A strong correlation between the turnover time and the periodicity of predominant oscillations has been demonstrated in cylinders with $\Gamma = 1$ \citep{Brown2009, Zuerner2019} or $\Gamma = 2$ \citep{Vogt2018b}. In these geometries, where a single-roll LSC exists, the oscillatory behaviour is due to torsion and sloshing at $\Gamma = 1$ or to JRV dynamics at $\Gamma = 2$, respectively. In our case of the $\Gamma = 5$ box, a torsional movement of single-roll LSC can be excluded. Nevertheless, it is reasonable to assume that perturbations of the flow pattern or instabilities of the boundary layers are transported at the characteristic velocity of circulation within the coherent structures. This interpretation was also adopted by \citet{Cioni1997} who assumed that the dominating frequency of the oscillations reflected in their temperature signal is associated with the velocity of the global circulation and path length of the LSC. This assumption leads to the relation $\frac{f_{OS}H^2}{\kappa} \varpropto RePr$ \citep{Cioni1997}. In the next section it becomes evident that this applies also for the flow configuration studied here, since the Reynolds number and the oscillation frequency follow the same power law as a function of $Ra$. 

\subsection{Transport of momentum and heat}
\label{subsec:momentum&heat}

\begin{figure}
\begin{center}
  \centerline{\includegraphics[width=0.5\linewidth]{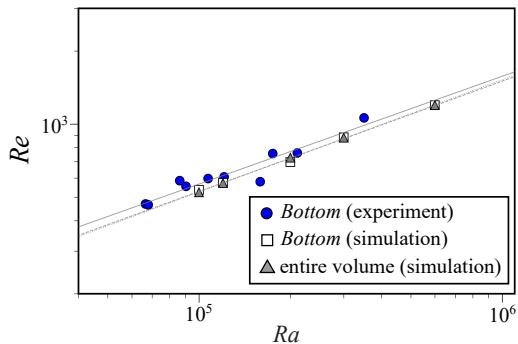}}
  \caption{Scaling of the Reynolds number $Re$ versus the Rayleigh number $Ra$: Results of numerical simulations are calculated for two characteristic velocities $u_{rms}$ and $U_{rms}$ representing the rms velocity for the measuring line of sensor $Bottom$ and the entire volume, respectively. Least-square approximation leads to $Re = 3.43Ra^{0.44}$ (grey solid lines) for the experimental data, $Re = 2.83Ra^{0.45}$ for numerical simulations (grey dashed line) and $Re = 2.56Ra^{0.46}$ for the entire volume (grey dotted line).}
\label{fig6}
\end{center}
\end{figure}

The Reynolds number $Re$ is the characteristic parameter to quantify the turbulent momentum transport in the convection cell. Our definition of $Re$ from the experiment relies on the root-mean-square (rms) velocity $u_{rms}$ determined from the linear velocity profile along the measuring line $Bottom$:

\begin{equation}
Re = \frac{u_{rms}H}{\nu}. 
\end{equation}

A simultaneous acquisition of all three velocity components in the entire container is difficult to achieve in the experiment.  With the limitation to one velocity component on one single measuring line, the $Re$ number derived from this appears relatively arbitrary and the question arises to what extent it is representative for the total flow and the momentum transfer. To check this, two $Re$ are determined by means of the numerical simulations, one corresponding to the experiment with $u_{rms}$ determined on the $Bottom$ measurement line and the other with the rms velocity $U_{rms}$ averaged over the entire volume in accordance with equation (\ref{Re_num}). To compensate for the difference that all three components of the velocity were considered in the calculation of $U_{rms}$  instead of only one in $u_{rms}$, the corresponding values of the $Re$ number have been corrected by a factor of $\sqrt{1/3}$.

The scaling behaviour of the resulting Reynolds numbers $Re$ is plotted in figure \ref{fig6}. There is no significant difference between the two results of numerical simulation, which suggests that the time-averaged velocities determined on the $Bottom$ measurement line are representative for the global flow. A power-law fit reveals a scaling law of $Re = 3.43Ra^{0.44}$ for velocity measurements and $Re = 2.83Ra^{0.45}$ for the numerical simulations. The direct comparison between experiment and simulation reveals a slight difference in the scaling exponents, but, the absolute values agree very well. 

A comparison with the previous study by \citet{Akashi2019} also reveals a slightly different scaling exponent, $u_{rms}(H/\kappa) = 0.059Ra^{0.50\pm 0.02}$. In \citet{Akashi2019} also smaller $Ra$, where unsteady roll arrangements are observed but no cellular flow regime occurs, were included into the power-law-fit. Thus, $Re$ in the cellular regime apparently shows a measurably more gradual increase with $Ra$ than in a roll regime. There are no differences between the current measurements and \citet{Akashi2019}, as long as only the range of the cellular regime is considered there.

Scaling exponents from $0.42$ to $0.53$ were reported by other experimental studies \citep{Takeshita1996, Cioni1997, Yanagisawa2013, Yanagisawa2015, Vogt2018b, Zuerner2019} and numerical simulations \citep{Scheel2017} for low $Pr$ thermal turbulence. \citet{Yanagisawa2013} found a higher velocity magnitude and increased scaling exponent, $u_{rms}(H/\kappa) = 0.08Ra^{0.53}$, in a square container with the same geometry filled with pure gallium corresponding to $Pr = 0.025$. \citet{Zuerner2019} defined three different characteristic velocities (1) the typical horizontal velocity magnitude near the plates $v_{LSC}$, (2) the typical vertical velocity magnitude of the LSC along the sidewall $v_{vert}$ and (3) the turbulent velocity fluctuations in the centre of the cell $v_{centre}$. This is reflected in slightly varying scaling exponents (1) $v_{LSC} \propto Ra^{0.42\pm 0.03}$, (2) $v_{vert}\propto Ra^{0.42\pm 0.04}$ and (3) $v_{centre}\propto Ra^{0.46\pm 0.04}$ \citep{Zuerner2019}. The smaller scaling exponent for the horizontal and vertical LSC is a result of probing different parts of the complex three-dimensional flow structure. It suggests the $Ra$ dependence of the representative velocities depends on how and where it is measured in the convection cell. It seems plausible that the option of calculating the rms velocity over the entire cell can be considered as the most reliable method for evaluating the representative convection velocity. For instance, \citet{Scheel2017} found a scaling, $U_{rms}\propto Ra^{0.45\pm 0.01}$ in a cylinder with $\Gamma = 1$ at $Pr = 0.021$, which match our results almost perfectly.  

\begin{figure}
\begin{center}
  \centerline{\includegraphics[width=0.5\linewidth]{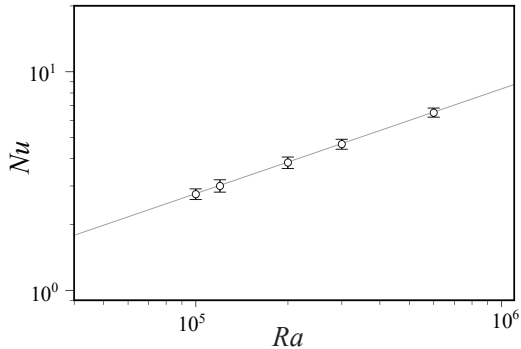}}
  \caption{Scaling of the Nusselt number $Nu$ versus $Ra$ obtained by numerical simulations, where the error bars represent the standard deviation of the temporal fluctuations and the grey line shows the fitted curve, $Nu = 0.14\times Ra^{0.26}$ (numerical simulation).}
\label{fig7}
\end{center}
\end{figure}

The heat transport of the three-dimensional cell structure is evaluated in numerical simulations by considering the Nusselt number dependence on the Rayleigh number, $Nu(Ra)$. Respective results are shown in figure \ref{fig7}. We find a scaling of $Nu= 0.14\times Ra^{0.26\pm0.01}$ which is in agreement to the outcome of measurements and simulations in a cylindrical cell with $\Gamma = 1$ \citep{Zuerner2019, Scheel2017}. The laboratory experiments conducted by \citet{Zuerner2019} for $Pr = 0.029$ provide $Nu\cong(0.12\pm0.04)\times Ra^{(0.27\pm0.04)}$ while the simulations by \citet{Scheel2017} suggest $Nu\cong(0.13\pm0.04)\times Ra^{(0.27\pm0.01)}$ for $Pr = 0.021$. Recent experiments conducted in the same container by \citet{Vogt2021} show a scaling of $Nu = 0.166\times Ra^{0.25}$ and thus a satisfactory agreement with the result presented here.

\section{Oscillatory dynamics of the three-dimensional cell structure}
\label{sec:dynamics}

\subsection{Procedure of phase averaging}
\label{subsec:proc-phas}

As shown by the flow measurements and the numerical simulations presented in the figures \ref{fig2}, \ref{fig2plus}, \ref{fig3}, respectively, the flow patterns are characterized by a variety of large and small-scale fluctuations, which indicate the presence of both a periodically oscillating coherent flow filling the entire container and inertia-dominated turbulence on smaller scales. We are interested in the dynamics of the coherent flow structure and want to separate it from the background of the turbulent fluctuations by a suitable averaging method. Investigations of turbulent superstructures in the very large aspect ratios are confronted with the same problem. \citet{Pandey2018} propose an averaging procedure where the averaging time to be used is in a range between the free-fall time $t_{ff}$ and the effective dissipation time $t_d = max(H^2/\nu, H^2/\kappa)$. In the case of turbulent superstructures in the very large aspect ratios, the patterns are slowly changing. The structure under consideration here undergoes significant oscillations with a constant periodicity. For this reason it is useful to apply a phase averaging to our numerical data. Such a procedure has already been suggested by \citet{Vogt2018b} to reveal the nature of the dynamic flow structures of the periodic oscillations of velocity and temperature in the $\Gamma = 2$ cylinder. The analysis covers one complete oscillation period $\tau_{OS} = 1/f_{OS}$, which in our case also corresponds to the turnover time for a respective fluid parcel (see figure \ref{fig5}).
It is based on simulation results with a computation time comprising 16 oscillation periods with a total number of 5536 snapshots. The entire data set is divided into 16 parts, each corresponding exactly to one oscillation period. This time period of one oscillation is again divided into 16 equidistant intervals, also called phases. Each interval or phase includes 21 snapshots of temperature, velocity and pressure. In the next step, these snapshots are averaged. Finally, for the distributions of temperature, velocity and pressure obtained for all oscillation periods are conditionally averaged with the respective distributions belonging to the same phase.

\citet{Vogt2018b} also had to adequately consider the orientation of the LSC in the convection cell for each time step, since the LSC symmetry plane in the cylindrical $\Gamma = 2$ cylinder wanders in the azimuthal direction. This did not have to be taken into account in our case, since the flow is approximately fixed in its orientation by the square geometry of the convection cell.

\subsection{Variations of temperature and flow field during the oscillation period}
\label{subsec:temp}

\begin{figure}
\begin{center}
  \centerline{\includegraphics[width=\linewidth]{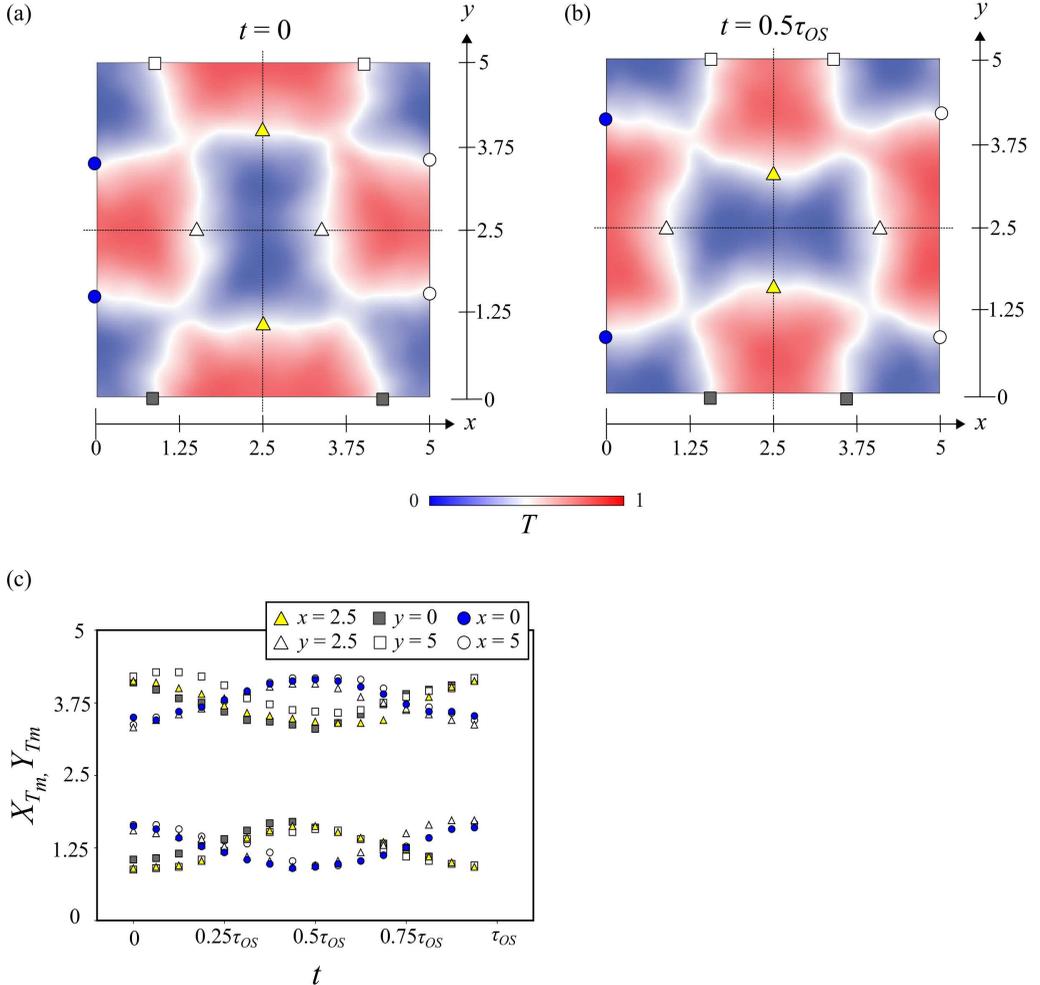}}
  \caption{Phase averaged temperature distribution for $Ra = 1.2\times10^5$ in the centre plane at (a) $t = 0$, (b) $t = 0.5\tau_{OS}$ and (c) time variations of the $x$ and $y$ positions of the mean temperature of $T_m = 0.5$ on the centre lines and the sidewalls of the container in the temperature distribution shown in (a) and (b)(numerical simulation). }
\label{fig8}
\end{center}
\end{figure}	

\begin{figure}
\begin{center}
  \centerline{\includegraphics[width=\linewidth]{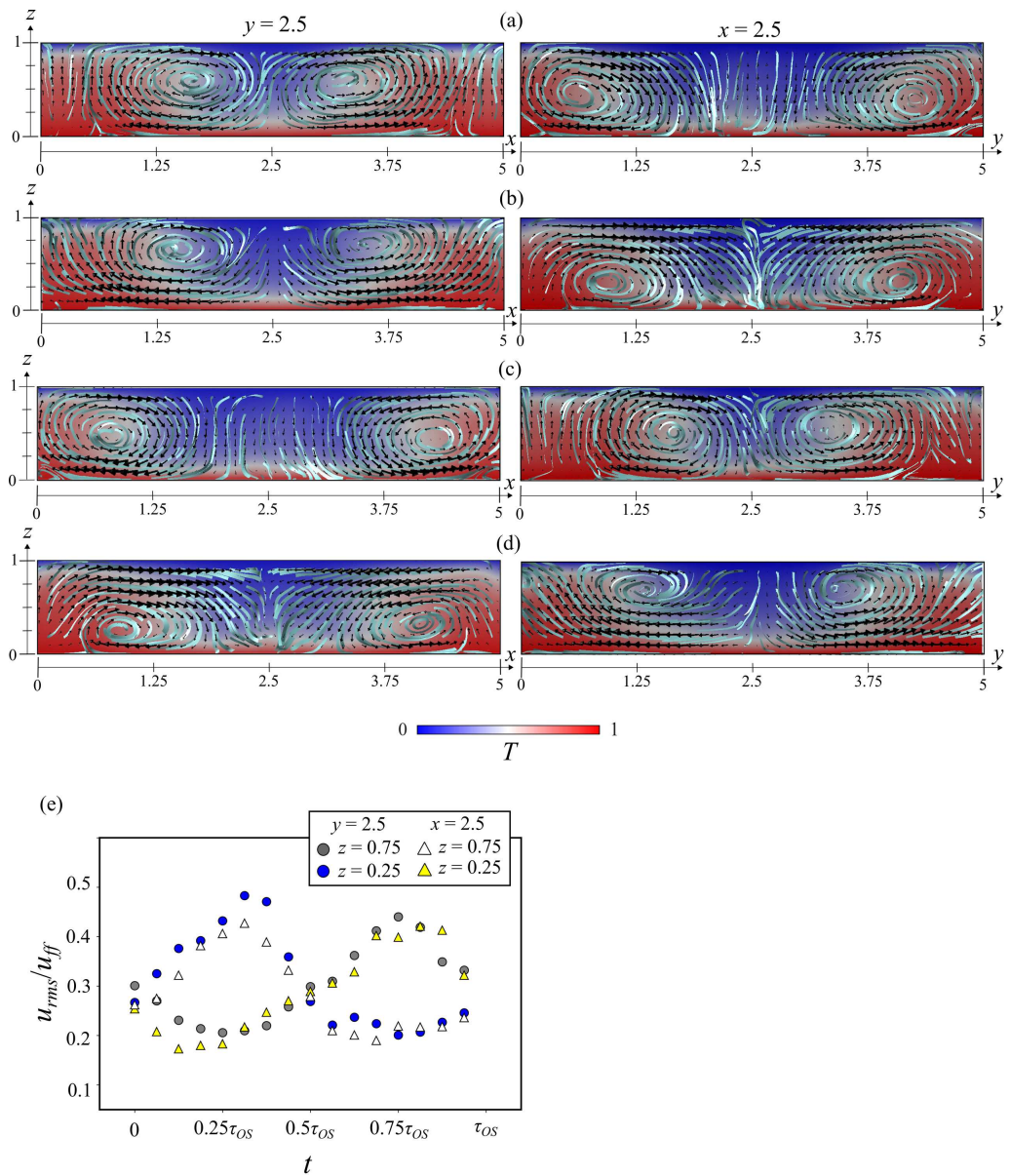}}
  \caption{Phase averaged temperature distribution with velocity vectors and stream lines for $Ra = 1.2\times10^5$ in the cross section of the centre lines of $y = 2.5$ (left column) and $x = 2.5$ (right column) at different phases:(a) $t=0$, (b) $t = 0.25\tau_{OS}$, (c) $t = 0.5\tau_{OS}$, and (d) $t = 0.75\tau_{OS}$. (e) Temporal change of representative velocity $u_{rms}$ calculated along the measurement lines of $z = 0.25$ and $z = 0.75$ in the cross sections of $y = 2.5$ and $x = 2.5$ shown in (a-d) (numerical simulation). }
\label{fig9}
\end{center}
\end{figure}	

\begin{figure}
\begin{center}
  \centerline{\includegraphics[width=\linewidth]{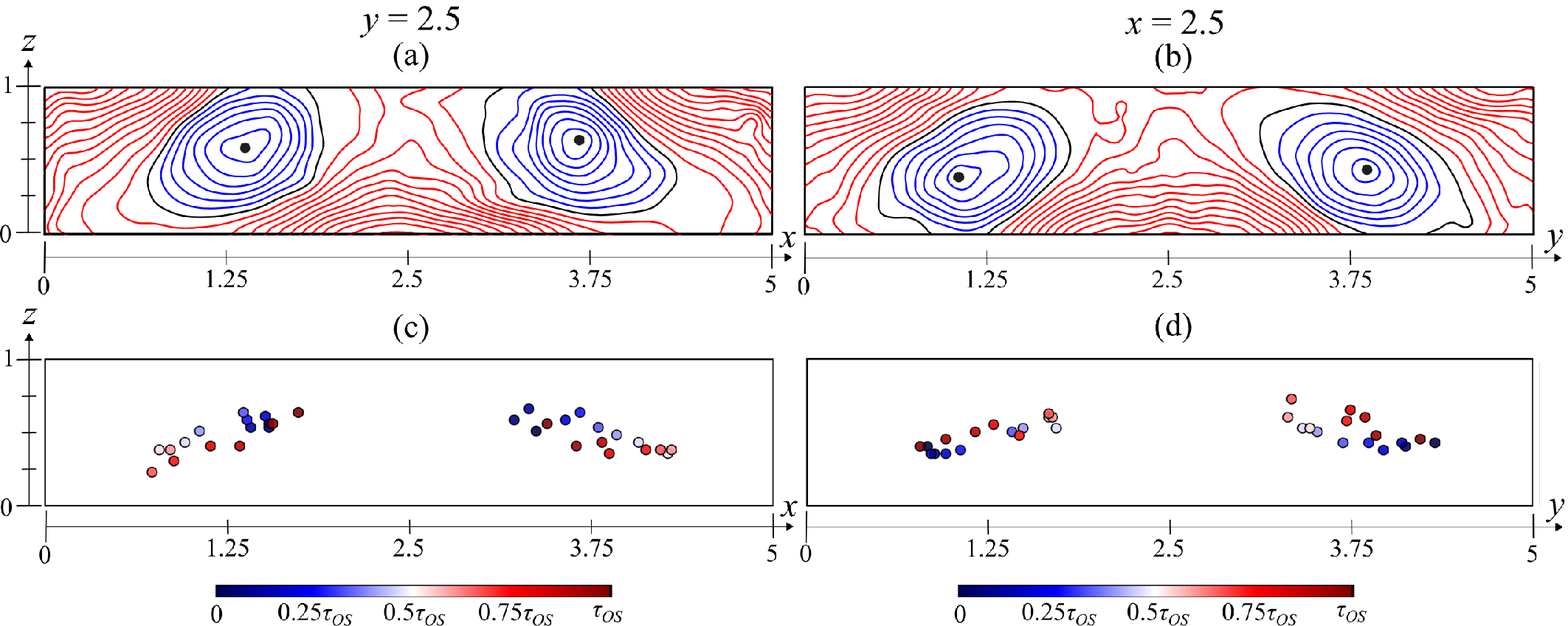}}
  \caption{Phase averaged pressure contour for $Ra = 1.2\times10^5$ in the cross section of (a) $y = 2.5$ (left column) and (b) $x = 2.5$ (right column) at the phase of $t = 0.25\tau_{OS}$. The minimum pressure points in the phase-averaged pressure contour during one oscillation period in the cross sections of (c) $y = 2.5$  and (d) $x = 2.5$ (numerical simulation). }
\label{fig10}
\end{center}
\end{figure}

The results presented in this chapter are restricted to $Ra = 1.2\times10^5$ as an example. The flow at other $Ra$ shows a qualitatively similar behaviour. The numerical calculations of temperature and flow fields are performed for the duration of 16 oscillations which corresponds to about 300 free fall times $t_{ff}$ and 5 thermal diffusion times $t_\kappa = H^2/\kappa$ with a grid resolution of $n_x  \times n_y \times n_z = 400 \times 400 \times 80$. Figures \ref{fig8}(a) and (b) show the phase averaged temperature distribution for the horizontal plane at the mid-height of $z = 0.5$ at two different phases. The temperature scale refers to the mean temperature $T_m = 0.5$ in the considered plane in such a way that the regions of the warm fluid marked by red colouration have a temperature above $T_m$, while colder fluid appears in blue colouration in the plots. We observe a distribution of temperature that resembles a checkerboard pattern, where the cold fluid occurs in the centre and the four corners of the container, while the warm fluid is located in centrally situated areas at the four sidewalls. This downwelling flow pattern is opposite from the upwelling one shown in the figure \ref{fig4}. The simulations reveal that the boundaries between the areas of warm and cold fluid which are identical with the isolines $T = T_m = 0.5$, are subject to significant deformation during the oscillations (See also supplementary material Movie 5). To quantify this behaviour in a simple way, we follow the displacements of some specially defined locations in the temperature distribution during one oscillation period. For this purpose we have selected the following positions: the intersection points of the $T_m$ isolines with the central lines of the  centre plane along the $x$-axis (white triangles) and the $y$-axis (yellow triangles) as well as at the intersection points of the $T_m$ isolines with the side walls. The latter are marked by the grey and white squares in $x$-direction and by the blue and white circles along the $y$-direction. Each of these marker points on the isolines can be assigned an $x$-coordinate $X_{T_m}$ and a $y$-coordinate $Y_{T_m}$. In the following, the movements of these points along the $x$ and $y$ centre lines and along the side walls are recorded. Our analysis takes into account that all marker points have only one degree of freedom and can change their position either exclusively along the $x$- or $y$-direction. The two snapshots in the figures \ref{fig8}(a) and (b) show in a certain sense two extreme states in which the respective pairs of points have either a minimum or maximum distance to each other. The oscillation of the flow pattern is reflected accordingly in the coordinates of these markers. 

This is demonstrated in figure \ref{fig8}(c) which displays the varying locations of the marker points during one oscillation period. A sinusoidal behaviour is detected, where the coordinates of each point accomplish a full oscillation in one cycle. It is noticeable that the selected pairs of markers always move in opposite directions, this concerns the white and yellow triangles on the centre lines as well as the points on the differently aligned side edges. If, for example, one considers the associated deformation of the fluid regions of higher temperature, this means that the approximately rectangular fields alternately retract to the area in front of the side wall or extend into the central zone. For further consideration we have defined the phase shown in figure \ref{fig8}(a) as the starting point $t = 0$ of the oscillation period. Because of the symmetry of the configuration, both states in the figures \ref{fig8}(a) and (b) are quasi identical, since the marker points reach their extreme positions here. The distances between all marker points are nearly equal at the intermediate times $t = 0.25\tau_{OS}$ and $t = 0.75\tau_{OS}$.     

Figures \ref{fig9}(a-d) show the cross section of the phase averaged temperature distribution, streamlines and velocity vectors in two vertical centre planes of the container at $y = 2.5$ and $x = 2.5$, respectively. The colour map represents the temperature of the fluid. The length of the velocity vectors indicates the magnitude of the velocity. The streamlines and the velocity vectors visualize two large-scale circulations lined up side by side, where the fluid sinks in the centre of the cell and rises on the side walls. Size and location of the two dominating vortices are not constant but subject to permanent changes. The positions of the rotational axes of the vortices are not stationary, but move in the shown sectional planes. Naturally, this has significant consequences on the distribution of cold and warm fluid, which is reflected in the temperature colour plot (See also supplementary material Movie 6). The changes in the size of the respective zones of cold and warm fluids are clearly visible in the plots and confirm the dynamics already described in the context of figure \ref{fig8}. It should also be pointed out here that the temperature distribution does not only change in the centre of the convection box, but that significant alterations also occur near the bottom and the lid. We will focus on this aspect in more detail later in section \ref{subsec:heat}.

For further evaluation, we consider the temporal behaviour of the rms velocities $u_{rms}$ recorded along two horizontal lines at $z = 0.25$ and $z = 0.75$ in both cross sections. Time variations of $u_{rms}$ during one oscillation period are plotted as shown in figure \ref{fig9}(e). The velocities $u_{rms}$ monitored along the various horizontal lines reach their maximum (minimum) values at $t = 0.25\tau_{OS}$ or half period later at $t= 0.75\tau_{OS}$. As already seen in the evaluation of the temperature distribution in figure \ref{fig8}(c) a sinusoidal behaviour of the velocities is observed. Furthermore, it is immediately noticeable that the velocities at the different heights in one plane and the velocities at the same height in different planes vary in opposite directions. This corresponds in an excellent way with the velocity measurements in the experiment (see \citet{Akashi2019}). The periodic up and down of the velocity magnitude near the lid or bottom is related to an oscillatory shift of the centre of the dominating vortices in both vertical and horizontal directions. In the figures \ref{fig9}(a-d) it can be seen that smaller secondary vortices can form temporarily on the side walls if the axes of the larger vortices are located near the centre of the convection container. The vertical shift of the vortex positions lets the zone around the horizontal centre line alternately come into the effective range of the horizontal flow pointing towards the centre or towards the side walls, respectively. This explains the flow patterns with periodic reversals of flow direction found in the mid-height of the convection cell by both measurements and simulations (see figures \ref{fig2plus}(b) and (c) and \ref{fig3}(b)).  

\subsection{Delineation of the three-dimensional flow structure}

In the previous section, it was shown that the location and properties of dominant vortices filling the entire height of the fluid layer change significantly during a period of oscillation.The position of a vortex centre can be determined by its axis of rotation, which is characterized by minimum pressure. The phase averaged pressure distribution has been calculated for this purpose. Snapshots of the pressure fields in both vertical sectional planes are displayed in the figures \ref{fig10}(a) and (b) for $t = 0.25\tau_{OS}$. Here, the positions of the minimum pressure are marked with dark blue coloured circles. The trajectories of these vortex axes during one oscillation period are presented in the figures \ref{fig10}(c) and (d). In addition to the information regarding the position of the pressure minimum in the respective sectional plane, the colouring of the circles in this diagram also indicates the time when the vortex centre was at which position. The trajectories of the axes of the large-scale circulations form the central segment of a diagonal line running approximately between the lower corners of the convection cell and the centre of the upper plate, with the forward and return paths not exactly on the same line. It becomes also obvious that the two circulation centres always move in opposite directions in their respective sectional planes with an offset of half a period.     

\begin{figure}
\begin{center}
  \centerline{\includegraphics[width=\linewidth]{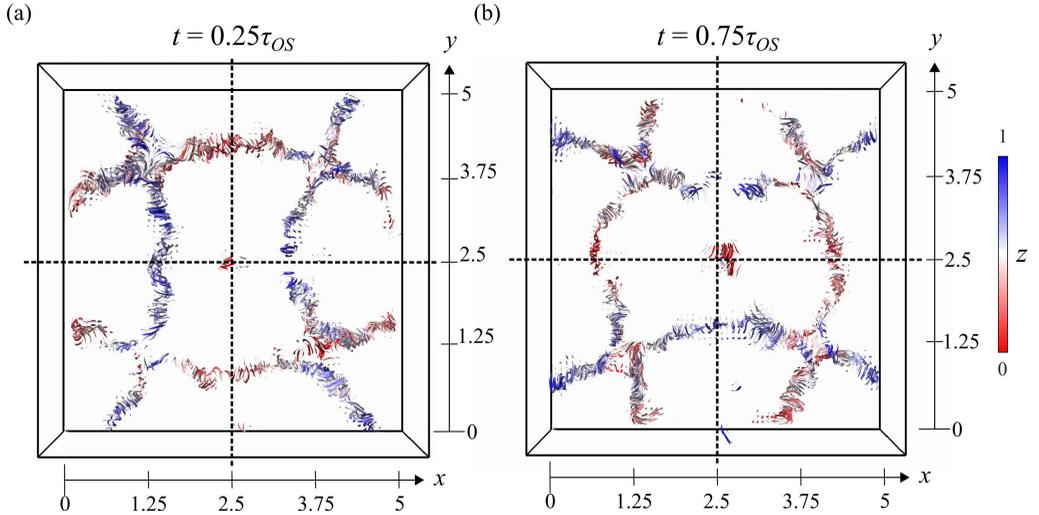}}
  \caption{Three-dimensional visualization of phase averaged stream lines at (a) $t = 0.25\tau_{OS}$ and (b) $t = 0.75\tau_{OS}$ for $Ra = 1.2\times10^5$. The dashed lines mark the two centre planes $x = 2.5$ and $y = 2.5$, which also represent the symmetry axes of the flow pattern. The colour represents the vertical position of the structure in the container (numerical simulation).}
\label{fig11}
\end{center}
\end{figure}
Figure \ref{fig11} is intended to illustrate the movement of the circulation areas in three dimensions by means of the streamlines in the vicinity of the vortex centres. To visualize the third component in vertical direction, the streamlines are coloured according to the vertical position of the structure. A first thing to notice is that four recirculating swirls exist in the convection cell. These four connected swirls are arranged as two pairs, with the axes of the respective pair aligned on average mainly along the $x$- and $y$-directions, respectively. Figures \ref{fig11}(a) and (b) contain two snapshots at $t = 0.25\tau_{OS}$ and $t = 0.75\tau_{OS}$, respectively. These are the moments when the swirls are either close to the bottom or to the top of the convection cell (see figures \ref{fig9} and \ref{fig10}). A corresponding animation of these results (supplementary material Movie 7-9) during one oscillation period shows, in fact, that we are dealing here with the phenomenon of a three-dimensional flow structure that shows the same characteristics as jump rope vortex (JRV), which was observed and described first by \citet{Vogt2018b}. The main difference to the geometry of a $\Gamma = 2$ cylinder, in which the dynamics of a JRV extend over almost the entire convection cell, is that four such structures exist in our $\Gamma = 5$ configuration, whose movements are restricted to each partial area of the fluid layer. Supplementary Movie 10 shows the phase-averaged temperature and velocity fields in the centre plane and the vertical cross sections of the limited area are in very good agreements with the conditionally averaged temperature and velocity fields in the centre plane, the LSC symmetry plane, and the perpendicular plane of the LSC symmetry plane in the cylinder of $\Gamma$ = 2 \citep{Vogt2018b}, respectively. It is interesting to note that the revolution of these swirls in the convection cell occurs in opposite direction to the main flow of the large-scale circulations. Another peculiarity is that the trajectories of the circulation centres obviously do not follow a circle, as one would normally expect for a typical jump rope, but through a tilted ellipse-like trajectory as visualized in figure \ref{fig10}(c) and (d). One reason for this might be the fact that the limited height of the fluid layer imposes a geometric constraint on the dynamics of the JRV. 

\subsection{Impact on the heat transport}
\label{subsec:heat}

The temperature distributions presented in section \ref{subsec:temp} demonstrate that considerable redistribution of cold and warm fluid occurs in the course of the three-dimensional oscillations. In this section, we consider the extent to which this has measurable effects on the heat transport. 

\begin{figure}
\begin{center}
  \centerline{\includegraphics[width=0.5\linewidth]{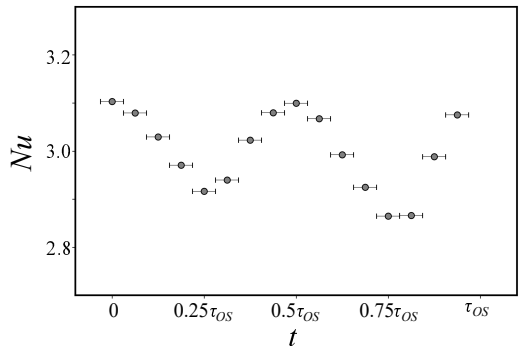}}
  \caption{Phase-averaged values of the Nusselt number $Nu$ during one oscillation period for for $Ra = 1.2\times10^5$ (numerical simulation).}
\label{fig12}
\end{center}
\end{figure}

\begin{figure}
\begin{center}
  \centerline{\includegraphics[width=\linewidth]{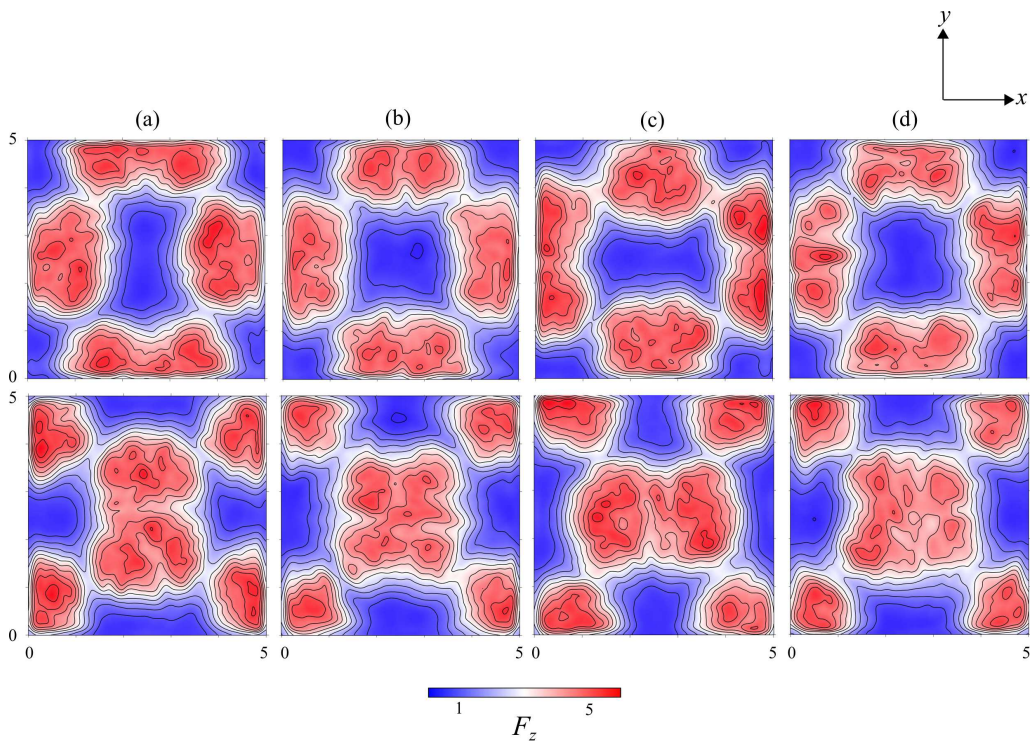}}
  \caption{Phase-averaged local heat flux at the top boundary, $F_{z, top}$ (top row), and at the bottom boundary, $F_{z, bot}$ (bottom row) for $Ra = 1.2\times10^5$, at different phases:(a) $t=0$, (b) $t = 0.25\tau_{OS}$, (c) $t = 0.5\tau_{OS}$, and (d) $t = 0.75\tau_{OS}$. The contour interval is 0.5 (numerical simulation)}. 
\label{fig13}
\end{center}
\end{figure}

To verify the steadiness of the heat transport, Nusselt numbers are calculated from the phase averaged temperature distributions. Figure \ref{fig12} presents the phase-averaged $Nu$ numbers during one oscillation period. The results reveal that $Nu$ is not constant but rather shows an oscillatory behaviour with a verifiable difference between the maximum and minimum value. Moreover, at the first glance it is a very astonishing finding that the period of the $Nu$ oscillation is only half as long as that of the fluctuations in the temperature field or the horizontal velocities.

Figure \ref{fig13} show contour plots of phase-averaged local heat flux. They represent local heat flux calculated at the top boundary, $F_{z, top}=-\left (\frac{\partial T}{\partial z} \right)_{z=1}$, and the bottom boundary, $F_{z, bot}=-\left (\frac{\partial T}{\partial z} \right)_{z=0}$, respectively. Here, the average number of local heat flux for an entire plate means the Nusselt number of the convection. The red coloured zones feature the local heat flux that is above the mean value of the entire area and period ($Nu \approx 3.0$), while local heat flux below the mean value are marked in blue colour. Here, the contrasting behaviour of the local heat flux at the bottom and the top is obvious which is caused by the flow that either impinges or detaches in the corresponding areas of the plates. The local heat flux distributions change during the oscillation period and agrees to a certain degree with the temperature distributions presented in figure \ref{fig8}. According to figure \ref{fig12}, $Nu$ reaches a maximum value at time points $t = 0$ and $t = 0.5\tau_{OS}$, while minima occur at $t = 0.25\tau_{OS}$ and $t = 0.75\tau_{OS}$. The corresponding local heat flux distributions on the plates are geometrically similar. This is in agreement with the periodicity of the $Nu$ number oscillations.

\begin{figure}
\begin{center}
  \centerline{\includegraphics[width=\linewidth]{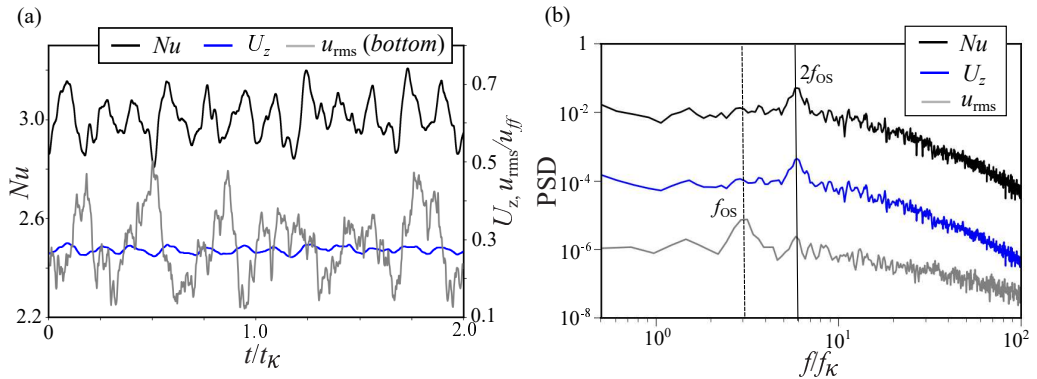}}
  \caption{(a) Time variations of $Nu$, the volume average of the vertical velocity component $U_z$, and the horizontal velocity $u_{rms}$ obtained by the line $Bottom$ at $z = 0.25$, and (b) the PSDs calculated from the temporal fluctuations of $Nu$, $U_z$ and the PSD calculated from the horizontal velocity $u_{rms}$ obtained by the line $Bottom$ for $Ra = 1.2\times10^5$ (numerical simulation). }
\label{fig15}
\end{center}
\end{figure}

The heat transfer of the system is mainly determined by the conductivity and stability of the thermal boundary layers \citep{Ahlers2009, Iyer2020}.
Our results presented in figure 10 suggest that the turbulent fluid motion might have a considerable impact on the
extent of the thermal boundary layers. Indeed, our data also show that the instantaneous thermal boundary layer
thickness at both the bottom and the top plates $\lambda_{\Theta}$ oscillates with the same frequency as the JRV. Owing to the relation $Nu = 1/(2\lambda_{\Theta})$ this is reflected
accordingly in the changes of $Nu$ illustrated in figure 13.

Figure \ref{fig15}(a) compares time series of $Nu$ with time series of the horizontal velocity $u_{rms}$ obtained along the line $Bottom$ and the volume-averaged vertical velocity for the entire container $U_{z}$ obtained by numerical simulations. All data show quasi-periodic oscillations. The signals of $Nu$ and $U_{z}$ are synchronized with each other, i.e. they oscillate with the same frequency, while $u_{rms}$ exhibits twice the oscillation period. This is confirmed by the corresponding PSDs presented in figure \ref{fig15}(b). Both PSDs of $Nu$ and $U_{z}$ show dominant peaks at the same frequency, which is almost exactly twice the oscillation frequency $f_{OS}$. Here we see an effect that, on closer inspection, is already apparent in figure \ref{fig9}. Figure \ref{fig9}(e) reveals that the approach of the centre of the circulation to one of the horizontal interfaces suppresses the horizontal flow there, while the horizontal flow on the opposite side simultaneously reaches a maximum. The evaluation of the numerical data in the whole convection cell shows that the vertical flow component, on the other hand, becomes greatest when the centre of the circulation moves either from bottom to top or in the opposite direction. This vertical momentum transport thus additionally contributes to the heat transport. Accordingly, figure \ref{fig15}(a) demonstrates a simultaneous occurrence of maximal values for $Nu$ and $U_z$.

\section{Summary and conclusions}

The present study combines flow measurements in a Rayleigh-Bénard setup and corresponding numerical simulations. The flow in the cuboid container with an aspect ratio of 5 is fully turbulent at a Prandtl number of 0.03 and Rayleigh numbers in the range of $Ra$ of $6.7\times10^4 \leq Ra \leq 3.5\times10^5$. In this parameter range the thermally driven convection produces a coherent cellular flow structure that is extremely stable with respect to changes in topology and orientation. On the other hand, this cellular flow pattern is subject to distinct oscillations. The change in fluid velocity during the oscillations is quite dramatic: the local velocity value varies during one period between a maximum value of the measured velocity and a value close to zero, in some regions even a periodic flow reversal occurs.

Our analysis succeeded in clarifying the complex 3D structure of the large-scale flow and identifying a multiple jump rope vortex (JRV) structure that is the underlying nature of the oscillations. The JRV structure was originally discovered by \citet{Vogt2018b} in a $\Gamma$ = 2 cylinder. As the name suggests, this is a vortex with a curved axis performing a cycling motion in three-dimensional space. Our results confirm the existence of such kind of flow structure also in a cuboid container with $\Gamma$ = 5. This finding is most remarkable, since it shows that the structure of the JRV is not a rare phenomenon that occurs only in special geometrical constellations. Instead, we showed that it is a more common property of turbulent convection. 

The striking feature of the structure detected here is that we find four vortices at the same time, with the symmetry axes of the circling motion being aligned parallel to the side walls of the fluid container in each case. One full revolution of the JRV corresponds exactly to one periodic length of the oscillation of the cellular flow structure. The dynamics of the JRV is strongly correlated and characterized by the fact that the respective opposite vortices move towards the centre or the side wall at the same time, while the other pair of vortices rotates with an offset of half a period. The JRVs interact significantly with the thermal boundary layers, its thickness decreases each time the centres of the JRVs are in close proximity in the immediate vicinity of the respective copper plate.  This also leaves detectable traces in the heat transfer in the form of an oscillation of the Nusselt number, its periodicity is related to the dynamics of the JRVs. Furthermore, instead of following almost circular path as observed in the $\Gamma$ = 2 cylinder, the vortices move along an elongated and obliquely inclined ellipse. Thus, it is obvious that the movement of the jump rope is affected by the limited height of the fluid layer.

The previous study by \citet{Akashi2019} revealed an increase in the characteristic length scale of the flow patterns with increasing $Ra$. In the slightly supercritical region directly after the onset of convection, roll structures are formed whose horizontal wavelength corresponds approximately to the height of the fluid layer $H$. As $Ra$ grows, the number of rolls decreases, thus the wavelength of the structures increases incrementally. Finally, there is a qualitative transition to the cellular regime considered here, the dimension of the cellular structure is $\thicksim 5H$. An increase in the characteristic length scale of coherent large-scale velocity structures with $Ra$ has also been reported in earlier studies, in particular for turbulent superstructures \citep{Fitzjarrald1976, Hartlep2003, Pandey2018, Krug2020}. Numerical studies identified the typical sizes of turbulent superstructures in large aspect ratios $\Gamma > 10$, which are in the order of 10$H$ for $Pr$ = 6.7 \citep{Busse1994, Pandey2018} or 6-7$H$ for $Pr$ = 0.7 \citep{Hartlep2003, Pandey2018, Stevens2018}. \citet{Sakievich2016, Sakievich2020} refer to these 'natural' dimensions of superstructures to explain the dominance of specific Fourier modes in a flat $\Gamma$ = 6.3 cylinder at $Pr$ = 6.7. From the data published by \citet{Pandey2018}, it is evident that the horizontal wave length of the superstructures in the range $Pr \leqq 10$ decreases when approaching smaller $Pr$ numbers. If we extrapolate the above estimates to the $Pr$ number of 0.03 considered here, we achieve approximately a size of 5$H$, a value that is found for the cellular regime. However, due to the limitation imposed by the finite $\Gamma$ = 5 box, it is not possible to verify whether the velocity field cannot form larger structures.

In summary, the cellular flow regime we studied here exhibits some general properties that are also attributed to turbulent superstructures \citep{Pandey2018,Stevens2018,Krug2020}:
\begin{itemize}
\item The characteristic length scale of the flow structures increases with increasing $Ra$ as shown by \citet{Akashi2019}.
\item The horizontal length scale is distinctly larger as the height of the flow domain. 
\item The characteristic dynamic time scale of the evolving or altering large-scale flow structure is much longer than the free-fall time.
\end{itemize}
In this general respect, our study is comparable to the work of \citet{Sakievich2016, Sakievich2020}, although the observed flow patterns differ from each other, which we believe is due to the differences in the shape of the convection vessels (cylinder vs. box). In particular, it is not surprising that the flow in the box is restricted in its dynamics and does not form a rotational symmetry as shown by \citet{Sakievich2020}. The cellular flow regime, instead, shows mirror symmetry with respect to the two vertical centre planes $x = 2.5$ and $y = 2.5$ of the convection box, respectively. This also reveals that the aspect ratios studied in both studies (Sakievichs' and ours) are still too small to produce flow structures that can develop unaffected by the sidewalls. New experiments with larger aspect ratios are needed to overcome this issue. Therefore, future work should consider further investigations in a wide parameter range of aspect ratios. In particular, it would be very interesting to find out to what extent the cellular structures studied in detail here also shape the flow in turbulent superstructures at large aspect ratios and whether such oscillatory dynamics with structures resembling JRV can also be observed there.

\section{Acknowledgements}
This work is supported by the Priority Programme SPP 1881 Turbulent Superstructures of the Deutsche Forschungsgemeinschaft (DFG) under the grant VO 2331/3. T.V. and F.S. also thank the DFG for the support under the grant VO 2331/1. The authors would like to thank Yuji Tasaka of Hokkaido University for fruitful discussions. Numerical simulations were performed on the Earth Simulator at JAMSTEC. \\

Declaration of Interests: The authors have no conflict of interest to declare.

\bibliographystyle{jfm}
\bibliography{JFM_Akashi}

\begin{thebibliography}{69}
\expandafter\ifx\csname natexlab\endcsname\relax\def\natexlab#1{#1}\fi
\def\au#1{#1} \def\ed#1{#1} \def\yr#1{#1}\def\at#1{#1}\def\jt#1{\textit{#1}}
  \def\bt#1{#1}\def\bvol#1{\textbf{#1}} \def\vol#1{#1} \def\pg#1{#1}
  \def\publ#1{#1}\def\arxiv#1{#1}\def\org#1{#1}\def\st#1{\textit{#1}}

\bibitem[Ahlers {\em et~al.\/}(2009)Ahlers, Grossmann \& Lohse]{Ahlers2009}
{\sc \au{Ahlers, G.}, \au{Grossmann, S.} \& \au{Lohse, D.}} \yr{2009}  \at{Heat
  transfer and large scale dynamics in turbulent {R}ayleigh-{B}\'enard
  convection}.  \jt{Rev. Mod. Phys.}  \bvol{81},  \pg{503--537}.

\bibitem[Akashi {\em et~al.\/}(2019)Akashi, Yanagisawa, Tasaka, Vogt, Murai \&
  Eckert]{Akashi2019}
{\sc \au{Akashi, M.}, \au{Yanagisawa, T.}, \au{Tasaka, Y.}, \au{Vogt, T.},
  \au{Murai, Y.} \& \au{Eckert, S.}} \yr{2019}  \at{Transition from convection
  rolls to large-scale cellular structures in turbulent {R}ayleigh-{B}\'enard
  convection in a liquid metal layer}.  \jt{Phys. Rev. Fluids}  \bvol{4},
  \pg{033501}.

\bibitem[Amati {\em et~al.\/}(2005)Amati, Koal, Massaioli, Sreenivasan \&
  Verzicco]{Amati2005}
{\sc \au{Amati, G.}, \au{Koal, K.}, \au{Massaioli, F.}, \au{Sreenivasan, K.~R.}
  \& \au{Verzicco, R.}} \yr{2005}  \at{Turbulent thermal convection at high
  {R}ayleigh numbers for a {B}oussinesq fluid of constant {P}randtl number}.
  \jt{Phys. Fluids}  \bvol{17}~(12),  \pg{121701}.

\bibitem[Bai {\em et~al.\/}(2016)Bai, Ji \& Brown]{Bai2016}
{\sc \au{Bai, K.}, \au{Ji, D.} \& \au{Brown, E.}} \yr{2016}  \at{Ability of a
  low-dimensional model to predict geometry-dependent dynamics of large-scale
  coherent structures in turbulence}.  \jt{Phys. Rev. E}  \bvol{93},
  \pg{023117}.

\bibitem[Bailon-Cuba {\em et~al.\/}(2010)Bailon-Cuba, Emran \&
  Schumacher]{Bailon2010}
{\sc \au{Bailon-Cuba, J.}, \au{Emran, M.~S.} \& \au{Schumacher, J.}} \yr{2010}
  \at{Aspect ratio dependence of heat transfer and large-scale flow in
  turbulent convection}.  \jt{J. Fluid Mech.}  \bvol{655},  \pg{152–173}.

\bibitem[Bénard(1900)]{Benard1900}
{\sc \au{Bénard, H.}} \yr{1900}  \at{Les tourbillons cellulaires dans une
  nappe liquide}.  \jt{Rev. Gén. Sci. Pures Appl.}  \bvol{11},
  \pg{1261–1276}.

\bibitem[Breuer {\em et~al.\/}(2004)Breuer, Wessling, Schmalzl \&
  Hansen]{Breuer2004}
{\sc \au{Breuer, M.}, \au{Wessling, S.}, \au{Schmalzl, J.} \& \au{Hansen, U.}}
  \yr{2004}  \at{Effect of inertia in {R}ayleigh–{B}énard convection}.
  \jt{Phys. Rev. E}  \bvol{69},  \pg{026302}.

\bibitem[Brown \& Ahlers(2009)]{Brown2009}
{\sc \au{Brown, E.} \& \au{Ahlers, G.}} \yr{2009}  \at{The origin of
  oscillations of the large-scale circulation of turbulent
  {R}ayleigh–{B}énard convection}.  \jt{J. Fluid Mech.}  \bvol{638},
  \pg{383–400}.

\bibitem[Brown {\em et~al.\/}(2005)Brown, Nikolaenko \& Ahlers]{Brown2005}
{\sc \au{Brown, E.}, \au{Nikolaenko, A.} \& \au{Ahlers, G.}} \yr{2005}
  \at{Reorientation of the large-scale circulation in turbulent
  {R}ayleigh–{B}énard convection}.  \jt{Phys. Rev. Lett.}  \bvol{95},
  \pg{084503}.

\bibitem[Burr \& Müller(2001)]{Burr2001}
{\sc \au{Burr, U.} \& \au{Müller, U.}} \yr{2001}  \at{{R}ayleigh–{B}énard
  convection in liquid metal layers under the influence of a vertical magnetic
  field}.  \jt{Phys. Fluids}  \bvol{13}~(11),  \pg{3247--3257}.

\bibitem[Busse(1994)]{Busse1994}
{\sc \au{Busse, F.H.}} \yr{1994}  \at{Spoke pattern convection}.  \jt{Acta
  Mechanica}  \bvol{4},  \pg{11--17}.

\bibitem[Busse {\em et~al.\/}(2003)Busse, Zaks \& Brausch]{Busse2003}
{\sc \au{Busse, F.H.}, \au{Zaks, M.A.} \& \au{Brausch, O.}} \yr{2003}
  \at{Centrifugally driven thermal convection at high {P}randtl numbers}.
  \jt{Physica D.}  \bvol{184}~(1),  \pg{3--20}.

\bibitem[Castaing {\em et~al.\/}(1989)Castaing, Gunaratne, Heslot, Kadanoff,
  Libchaber, Thomae, Wu, Zaleski \& Zanetti]{Castaing1989}
{\sc \au{Castaing, B.}, \au{Gunaratne, G.}, \au{Heslot, F.}, \au{Kadanoff, L.},
  \au{Libchaber, A.}, \au{Thomae, S.}, \au{Wu, X.}, \au{Zaleski, S.} \&
  \au{Zanetti, G.}} \yr{1989}  \at{Scaling of hard thermal turbulence in
  {R}ayleigh–{B}énard convection}.  \jt{J. Fluid Mech.}  \bvol{204},
  \pg{1–30}.

\bibitem[Chandrasekhar(1961)]{Chandrasekhar1961}
{\sc \au{Chandrasekhar, S.}} \yr{1961} {\em Hydrodynamic and Hydromagnetic
  Stability\/}.  \publ{Dover, New York}.

\bibitem[Chillà \& Schumacher(2012)]{Chilla2012}
{\sc \au{Chillà, F.} \& \au{Schumacher, J.}} \yr{2012}  \at{New perspectives
  in turbulent {R}ayleigh-{B}énard convection}.  \jt{Eur. Phys. J. E.}
  \bvol{35(7)},  \pg{58}.

\bibitem[Chorin(1967)]{Chorin1967}
{\sc \au{Chorin, A.~J.}} \yr{1967}  \at{A numerical method for solving
  incompressible viscous flow problems}.  \jt{J. Comput. Phys.}  \bvol{2}~(1),
  \pg{12 -- 26}.

\bibitem[Cioni {\em et~al.\/}(1997)Cioni, Cilberto \& Sommeria]{Cioni1997}
{\sc \au{Cioni, S.}, \au{Cilberto, S.} \& \au{Sommeria, J.}} \yr{1997}
  \at{Strongly turbulent {R}ayleigh–{B}énard convection in mercury:
  comparison with results at moderate {P}randtl number}.  \jt{J. Fluid. Mech.}
  \bvol{335},  \pg{111–140}.

\bibitem[Cramer {\em et~al.\/}(2004)Cramer, Zhang \& Eckert]{Cramer2004}
{\sc \au{Cramer, A.}, \au{Zhang, C.} \& \au{Eckert, S.}} \yr{2004}  \at{Local
  flow structures in liquid metals measured by ultrasonic {D}oppler
  velocimetry}.  \jt{Flow Meas. Instrum.}  \bvol{15},  \pg{145--153}.

\bibitem[Eckert \& Gerbeth(2002)]{Eckert2002}
{\sc \au{Eckert, S.} \& \au{Gerbeth, G.}} \yr{2002}  \at{Velocity measurements
  in liquid sodium by means of ultrasound {D}oppler velocimetry}.  \jt{Exp.
  Fluids}  \bvol{32},  \pg{542--546}.

\bibitem[Emran \& Schumacher(2015)]{Emran2015}
{\sc \au{Emran, M.~S.} \& \au{Schumacher, J.}} \yr{2015}  \at{Large-scale mean
  patterns in turbulent convection}.  \jt{J. Fluid. Mech.}  \bvol{776},
  \pg{96–108}.

\bibitem[Fitzjarrald(1976)]{Fitzjarrald1976}
{\sc \au{Fitzjarrald, D.E.}} \yr{1976}  \at{An experimental study of turbulent
  convection in air}.  \jt{J. Fluid. Mech.}  \bvol{73},  \pg{693--719}.

\bibitem[Foroozani {\em et~al.\/}(2017)Foroozani, Niemela, Armenio \&
  Sreenivasan]{Foroozani2017}
{\sc \au{Foroozani, N.}, \au{Niemela, J.~J.}, \au{Armenio, V.} \&
  \au{Sreenivasan, K.~R.}} \yr{2017}  \at{Reorientations of the large-scale
  flow in turbulent convection in a cube}.  \jt{Phys. Rev. E}  \bvol{95},
  \pg{033107}.

\bibitem[Funfschilling \& Ahlers(2004)]{Funfschilling2004}
{\sc \au{Funfschilling, D.} \& \au{Ahlers, G.}} \yr{2004}  \at{Plume motion and
  large-scale circulation in a cylindrical {R}ayleigh-{B}\'enard cell}.
  \jt{Phys. Rev. Lett.}  \bvol{92},  \pg{194502}.

\bibitem[Grossmann \& Lohse(2000)]{Grossmann2000}
{\sc \au{Grossmann, S.} \& \au{Lohse, D.}} \yr{2000}  \at{Scaling in thermal
  convection: a unifying theory}.  \jt{J. Fluid. Mech.}  \bvol{407},
  \pg{27--56}.

\bibitem[Grossmann \& Lohse(2002)]{Grossmann2002}
{\sc \au{Grossmann, S.} \& \au{Lohse, D.}} \yr{2002}  \at{Prandtl and
  {R}ayleigh number dependence of the {R}eynolds number in turbulent thermal
  convection}.  \jt{Phys. Rev. E}  \bvol{66},  \pg{016305}.

\bibitem[von Hardenberg {\em et~al.\/}(2008)von Hardenberg, Parodi, Passoni,
  Provenzale \& Spiegel]{VonHardenberg2008}
{\sc \au{von Hardenberg, J.}, \au{Parodi, A.}, \au{Passoni, G.},
  \au{Provenzale, A.} \& \au{Spiegel, E.~A.}} \yr{2008}  \at{Large-scale
  patterns in {R}ayleigh–{B}énard convection}.  \jt{Physics Letters A}
  \bvol{372}~(13),  \pg{2223--2229}.

\bibitem[Hartlep {\em et~al.\/}(2003)Hartlep, Tilgner \& Busse]{Hartlep2003}
{\sc \au{Hartlep, T.}, \au{Tilgner, A.} \& \au{Busse, F.~H.}} \yr{2003}
  \at{Large scale structures in {R}ayleigh-{B}\'enard convection at high
  {R}ayleigh numbers}.  \jt{Phys. Rev. Lett.}  \bvol{91},  \pg{064501}.

\bibitem[Horanyi {\em et~al.\/}(1999)Horanyi, Krebs \& Müller]{Horanyi1999}
{\sc \au{Horanyi, S.}, \au{Krebs, L.} \& \au{Müller, U.}} \yr{1999}
  \at{Turbulent {R}ayleigh–{B}énard convection in low {P}randtl–number
  fluids}.  \jt{Int. J. Heat Mass Transf.}  \bvol{42}~(21),  \pg{3983 -- 4003}.

\bibitem[Iyer {\em et~al.\/}(2020)Iyer, Scheel, Schumacher \&
  Sreenivasan]{Iyer2020}
{\sc \au{Iyer, K.~P.}, \au{Scheel, J.~D.}, \au{Schumacher, J.} \&
  \au{Sreenivasan, K.~R.}} \yr{2020}  \at{Classical 1/3 scaling of convection
  holds up to {R}a = 10{$^{15}$}}.  \jt{Proc. Natl Acad. Sci. USA}
  \bvol{117}~(14),  \pg{7594--7598}.

\bibitem[Krishnamurti \& Howard(1981)]{Krishnamurti1981}
{\sc \au{Krishnamurti, R.} \& \au{Howard, L.~N.}} \yr{1981}  \at{Large-scale
  flow generation in turbulent convection}.  \jt{Proc. Natl Acad. Sci. USA.}
  \bvol{78}~(4),  \pg{1981--1985}.

\bibitem[Krug {\em et~al.\/}(2020)Krug, Lohse \& Stevens]{Krug2020}
{\sc \au{Krug, D.}, \au{Lohse, D.} \& \au{Stevens, R. J. A.~M.}} \yr{2020}
  \at{Coherence of temperature and velocity superstructures in turbulent
  {R}ayleigh–{B}énard flow}.  \jt{J. Fluid. Mech.}  \bvol{887},  \pg{A2}.

\bibitem[Mashiko {\em et~al.\/}(2004)Mashiko, Tsuji, Mizuno \&
  Sano]{Mashiko2004}
{\sc \au{Mashiko, T.}, \au{Tsuji, Y.}, \au{Mizuno, T.} \& \au{Sano, M.}}
  \yr{2004}  \at{Instantaneous measurement of velocity fields in developed
  thermal turbulence in mercury}.  \jt{Phys. Rev. E}  \bvol{69},  \pg{036306}.

\bibitem[Niemela {\em et~al.\/}(2001)Niemela, Skrbek, Sreenivasan \&
  Donnely]{Niemela2001}
{\sc \au{Niemela, J.~J.}, \au{Skrbek, L.}, \au{Sreenivasan, K.~R.} \&
  \au{Donnely, R.~J.}} \yr{2001}  \at{The wind in confined thermal convection}.
   \jt{J. Fluid. Mech.}  \bvol{449},  \pg{169–178}.

\bibitem[Pandey {\em et~al.\/}(2018)Pandey, Scheel \& Schumacher]{Pandey2018}
{\sc \au{Pandey, A.}, \au{Scheel, J.~D.} \& \au{Schumacher, J.}} \yr{2018}
  \at{Turbulent superstructures in {R}ayleigh-{B}énard convection}.  \jt{Nat.
  Commun.}  \bvol{9}.

\bibitem[Plevachuk {\em et~al.\/}(2014)Plevachuk, Sklyarchuk, Eckert, Gerbeth
  \& Novakovic]{Plevachuk2014}
{\sc \au{Plevachuk, Y.}, \au{Sklyarchuk, V.}, \au{Eckert, S.}, \au{Gerbeth, G.}
  \& \au{Novakovic, R.}} \yr{2014}  \at{Thermophysical properties of the liquid
  {G}a–{I}n–{S}n eutectic alloy}.  \jt{J. Chem. Eng. Data}  \bvol{59},
  \pg{757–763}.

\bibitem[du~Puits {\em et~al.\/}(2007)du~Puits, Resagk \& Thess]{duPuits2007}
{\sc \au{du~Puits, R.}, \au{Resagk, C.} \& \au{Thess, A.}} \yr{2007}
  \at{Breakdown of wind in turbulent thermal convection}.  \jt{Phys. Rev. E}
  \bvol{75},  \pg{016302}.

\bibitem[Qiu \& Tong(2001)]{Qiu2001}
{\sc \au{Qiu, X.-L.} \& \au{Tong, P.}} \yr{2001}  \at{Onset of coherent
  oscillations in turbulent {R}ayleigh-{B}énard convection}.  \jt{Phys. Rev.
  Lett.}  \bvol{87},  \pg{094501}.

\bibitem[Sakievich {\em et~al.\/}(2016)Sakievich, Peet \&
  Adrian]{Sakievich2016}
{\sc \au{Sakievich, P.~J.}, \au{Peet, Y.~T.} \& \au{Adrian, R.~J.}} \yr{2016}
  \at{Large-scale thermal motions of turbulent {R}ayleigh–{B}énard
  convection in a wide aspect-ratio cylindrical domain}.  \jt{Int. J. Heat
  Fluid Flow}  \bvol{61},  \pg{183--196}.

\bibitem[Sakievich {\em et~al.\/}(2020)Sakievich, Peet \&
  Adrian]{Sakievich2020}
{\sc \au{Sakievich, P.~J.}, \au{Peet, Y.~T.} \& \au{Adrian, R.~J.}} \yr{2020}
  \at{Temporal dynamics of large-scale structures for turbulent
  {R}ayleigh–{B}énard convection in a moderate aspect-ratio cylinder}.
  \jt{J. Fluid. Mech}  \bvol{901}.

\bibitem[Scheel {\em et~al.\/}(2013)Scheel, Emran \& Schumacher]{Scheel2013}
{\sc \au{Scheel, J.~D}, \au{Emran, M.~S} \& \au{Schumacher, J.}} \yr{2013}
  \at{Resolving the fine-scale structure in turbulent
  {R}ayleigh{\textendash}b{\'{e}}nard convection}.  \jt{New J. Phys.}
  \bvol{15},  \pg{113063}.

\bibitem[Scheel \& Schumacher(2017)]{Scheel2017}
{\sc \au{Scheel, J.~D.} \& \au{Schumacher, J.}} \yr{2017}  \at{Predicting
  transition ranges to fully turbulent viscous boundary layers in low {P}randtl
  number convection flows}.  \jt{Phys. Rev. Fluids}  \bvol{2},  \pg{123501}.

\bibitem[Schneide {\em et~al.\/}(2018)Schneide, Pandey, Padberg-Gehle \&
  Schumacher]{Schneide2018}
{\sc \au{Schneide, C.}, \au{Pandey, A.}, \au{Padberg-Gehle, K.} \&
  \au{Schumacher, J.}} \yr{2018}  \at{Probing turbulent superstructures in
  {R}ayleigh-{B}\'enard convection by {L}agrangian trajectory clusters}.
  \jt{Phys. Rev. Fluids}  \bvol{3},  \pg{113501}.

\bibitem[Schumacher \& Scheel(2016)]{Schumacher2016}
{\sc \au{Schumacher, J.} \& \au{Scheel, J.~D.}} \yr{2016}  \at{Extreme
  dissipation event due to plume collision in a turbulent convection cell}.
  \jt{Phys. Rev. E}  \bvol{94},  \pg{043104}.

\bibitem[Segawa {\em et~al.\/}(1998)Segawa, Naert \& Sano]{Segawa1998}
{\sc \au{Segawa, T.}, \au{Naert, A.} \& \au{Sano, M.}} \yr{1998}  \at{Matched
  boundary layers in turbulent {R}ayleigh-{B}\'enard convection of mercury}.
  \jt{Phys. Rev. E}  \bvol{57},  \pg{557--560}.

\bibitem[Siggia(1994)]{Siggia1994}
{\sc \au{Siggia, E.D.}} \yr{1994}  \at{High {R}ayleigh number convection}.
  \jt{Ann. Rev. Fluid Mech.}  \bvol{26},  \pg{137--168}.

\bibitem[Stevens {\em et~al.\/}(2011)Stevens, Clercx \& Lohse]{Stevens2011}
{\sc \au{Stevens, R.J. A.~M.}, \au{Clercx, H. J.~H.} \& \au{Lohse, D.}}
  \yr{2011}  \at{Effect of plumes on measuring the large scale circulation in
  turbulent {R}ayleigh-{B}énard convection}.  \jt{Phys. Fluids}
  \bvol{23}~(9),  \pg{095110}.

\bibitem[Stevens {\em et~al.\/}(2018)Stevens, Blass, Zhu, Verzicco \&
  Lohse]{Stevens2018}
{\sc \au{Stevens, R. J. A.~M.}, \au{Blass, A.}, \au{Zhu, X.}, \au{Verzicco, R.}
  \& \au{Lohse, D.}} \yr{2018}  \at{Turbulent thermal superstructures in
  {R}ayleigh-{B}\'enard convection}.  \jt{Phys. Rev. Fluids}  \bvol{3},
  \pg{041501}.

\bibitem[Stevens {\em et~al.\/}(2013)Stevens, van~der Poel, Grossmann \&
  Lohse]{Stevens2013}
{\sc \au{Stevens, R. J. A.~M.}, \au{van~der Poel, E.~P.}, \au{Grossmann, S.} \&
  \au{Lohse, D.}} \yr{2013}  \at{The unifying theory of scaling in thermal
  convection: the updated prefactors}.  \jt{J. Fluid. Mech}  \bvol{730},
  \pg{295–308}.

\bibitem[Stringano \& Verzicco(2006)]{Stringano2006}
{\sc \au{Stringano, G.} \& \au{Verzicco, R.}} \yr{2006}  \at{Mean flow
  structure in thermal convection in a cylindrical cell of aspect ratio one
  half}.  \jt{J. Fluid. Mech.}  \bvol{548},  \pg{1–16}.

\bibitem[Sun {\em et~al.\/}(2005{\natexlab{{\em a\/}}})Sun, Xi \&
  Xia]{Sun2005b}
{\sc \au{Sun, C.}, \au{Xi, H.-D.} \& \au{Xia, K.-Q.}} \yr{2005{\natexlab{{\em
  a\/}}}}  \at{Azimuthal symmetry, flow dynamics, and heat transport in
  turbulent thermal convection in a cylinder with an aspect ratio of 0.5}.
  \jt{Phys. Rev. Lett.}  \bvol{95},  \pg{074502}.

\bibitem[Sun {\em et~al.\/}(2005{\natexlab{{\em b\/}}})Sun, Xia \&
  Tong]{Sun2005a}
{\sc \au{Sun, C.}, \au{Xia, K.~Q.} \& \au{Tong, P.}} \yr{2005{\natexlab{{\em
  b\/}}}}  \at{Three-dimensional flow structures and dynamics of turbulent
  thermal convection in a cylindrical cell}.  \jt{Phys. Rev. E}  \bvol{72},
  \pg{026302}.

\bibitem[Takeda(1987)]{Takeda1987}
{\sc \au{Takeda, Y.}} \yr{1987}  \at{Measurement of velocity profile of mercury
  flow by ultrasound doppler shift method}.  \jt{Nucl. Technol.}
  \bvol{79}~(1),  \pg{120--124}.

\bibitem[Takeshita {\em et~al.\/}(1996)Takeshita, Segawa, Glazier \&
  Sano]{Takeshita1996}
{\sc \au{Takeshita, T.}, \au{Segawa, T.}, \au{Glazier, J.~A.} \& \au{Sano, M.}}
  \yr{1996}  \at{Thermal turbulence in mercury}.  \jt{Phys. Rev. Lett.}
  \bvol{76},  \pg{1465--1468}.

\bibitem[Tasaka {\em et~al.\/}(2016)Tasaka, Igaki, Yanagisawa, Vogt, Zuerner \&
  Eckert]{Tasaka2016}
{\sc \au{Tasaka, Y.}, \au{Igaki, K.}, \au{Yanagisawa, T.}, \au{Vogt, T.},
  \au{Zuerner, T.} \& \au{Eckert, S.}} \yr{2016}  \at{Regular flow reversals in
  {R}ayleigh-{B}\'enard convection in a horizontal magnetic field}.  \jt{Phys.
  Rev. E}  \bvol{93},  \pg{043109}.

\bibitem[Tasaka {\em et~al.\/}(2021)Tasaka, Yanagisawa, Fujita, Miyagoshi \&
  Sakuraba]{Tasaka2021}
{\sc \au{Tasaka, Y.}, \au{Yanagisawa, T.}, \au{Fujita, K.}, \au{Miyagoshi, T.}
  \& \au{Sakuraba, A.}} \yr{2021}  \at{Two-dimensional oscillation of
  convection roll in a finite liquid metal layer under a horizontal magnetic
  field}.  \jt{J. Fluid. Mech.}  \bvol{911},  \pg{A19}.

\bibitem[Tsuji {\em et~al.\/}(2005)Tsuji, Mizuno, Mashiko \& Sano]{Tsuji2005}
{\sc \au{Tsuji, Y.}, \au{Mizuno, T.}, \au{Mashiko, T.} \& \au{Sano, M.}}
  \yr{2005}  \at{Mean wind in convective turbulence of mercury}.  \jt{Phys.
  Rev. Lett.}  \bvol{94},  \pg{034501}.

\bibitem[Verzicco \& Camussi(2003)]{Verzicco2003}
{\sc \au{Verzicco, R.} \& \au{Camussi, R}} \yr{2003}  \at{Numerical experiments
  on strongly turbulent thermal convection in a slender cylindrical cell}.
  \jt{J. Fluid. Mech.}  \bvol{477},  \pg{19–49}.

\bibitem[Villermaux(1995)]{Villermaux1995}
{\sc \au{Villermaux, E.}} \yr{1995}  \at{Memory-induced low frequency
  oscillations in closed convection boxes}.  \jt{Phys. Rev. Lett.}  \bvol{75},
  \pg{4618--4621}.

\bibitem[Vogt {\em et~al.\/}(2018{\natexlab{{\em a\/}}})Vogt, Horn, Grannan \&
  Aurnou]{Vogt2018b}
{\sc \au{Vogt, T.}, \au{Horn, S.}, \au{Grannan, A.~M.} \& \au{Aurnou, J.~M.}}
  \yr{2018{\natexlab{{\em a\/}}}}  \at{Jump rope vortex in liquid metal
  convection}.  \jt{Proc. Natl Acad. Sci. USA}  \bvol{115}~(50),
  \pg{12674--12679}.

\bibitem[Vogt {\em et~al.\/}(2018{\natexlab{{\em b\/}}})Vogt, Ishimi,
  Yanagisawa, Tasaka, Sakuraba \& Eckert]{Vogt2018a}
{\sc \au{Vogt, T.}, \au{Ishimi, W.}, \au{Yanagisawa, T.}, \au{Tasaka, Y.},
  \au{Sakuraba, A.} \& \au{Eckert, S.}} \yr{2018{\natexlab{{\em b\/}}}}
  \at{Transition between quasi-two-dimensional and three-dimensional
  {R}ayleigh-{B}\'enard convection in a horizontal magnetic field}.  \jt{Phys.
  Rev. Fluids}  \bvol{3},  \pg{013503}.

\bibitem[Vogt {\em et~al.\/}(2021)Vogt, Yang, Schindler \& Eckert]{Vogt2021}
{\sc \au{Vogt, T.}, \au{Yang, J.~C.}, \au{Schindler, F.} \& \au{Eckert, S.}}
  \yr{2021}  \at{Free-fall velocities and heat transport enhancement in liquid
  metal magneto-convection}.  \jt{J. Fluid. Mech.}  \bvol{915},  \pg{A68}.

\bibitem[Xi \& Xia(2008)]{Xi2008}
{\sc \au{Xi, H.} \& \au{Xia, K.-Q.}} \yr{2008}  \at{Flow mode transitions in
  turbulent thermal convection}.  \jt{Phys. Fluids.}  \bvol{20}~(5),
  \pg{055104}.

\bibitem[Xi {\em et~al.\/}(2009)Xi, Zhou, Zhou, Chan \& Xia]{Xi2009}
{\sc \au{Xi, H.-D.}, \au{Zhou, S.-Q.}, \au{Zhou, Q.}, \au{Chan, T.-S.} \&
  \au{Xia, K.-Q.}} \yr{2009}  \at{Origin of the temperature oscillation in
  turbulent thermal convection}.  \jt{Phys. Rev. Lett.}  \bvol{102},
  \pg{044503}.

\bibitem[Yanagisawa {\em et~al.\/}(2013)Yanagisawa, Hamano, Miyagoshi,
  Yamagishi, Tasaka \& Takeda]{Yanagisawa2013}
{\sc \au{Yanagisawa, T.}, \au{Hamano, Y.}, \au{Miyagoshi, T.}, \au{Yamagishi,
  Y.}, \au{Tasaka, Y.} \& \au{Takeda, Y.}} \yr{2013}  \at{Convection patterns
  in a liquid metal under an imposed horizontal magnetic field}.  \jt{Phys.
  Rev. E}  \bvol{88},  \pg{063020}.

\bibitem[Yanagisawa {\em et~al.\/}(2015)Yanagisawa, Hamano \&
  Sakuraba]{Yanagisawa2015}
{\sc \au{Yanagisawa, T.}, \au{Hamano, Y.} \& \au{Sakuraba, A.}} \yr{2015}
  \at{Flow reversals in low-{P}randtl-number {R}ayleigh-{B}\'enard convection
  controlled by horizontal circulations}.  \jt{Phys. Rev. E}  \bvol{92},
  \pg{023018}.

\bibitem[Yanagisawa {\em et~al.\/}(2010)Yanagisawa, Yamagishi, Hamano, Tasaka,
  Yoshida, Yano \& Takeda]{Yanagisawa2010}
{\sc \au{Yanagisawa, T.}, \au{Yamagishi, Y.}, \au{Hamano, Y.}, \au{Tasaka, Y.},
  \au{Yoshida, M.}, \au{Yano, K.} \& \au{Takeda, Y.}} \yr{2010}  \at{Structure
  of large-scale flows and their oscillation in the thermal convection of
  liquid gallium}.  \jt{Phys. Rev. E}  \bvol{82},  \pg{016320}.

\bibitem[Yang {\em et~al.\/}(2021)Yang, Vogt \& Eckert]{Yang2021}
{\sc \au{Yang, J.~C.}, \au{Vogt, T.} \& \au{Eckert, S.}} \yr{2021}
  \at{Transition from steady to oscillating convection rolls in
  {R}ayleigh-{B}\'enard convection under the influence of a horizontal magnetic
  field}.  \jt{Phys. Rev. Fluids}  \bvol{6},  \pg{023502}.

\bibitem[Zhou {\em et~al.\/}(2009)Zhou, Xi, Zhou, Sun \& Xia]{Zhou2009}
{\sc \au{Zhou, Q.}, \au{Xi, H.-D.}, \au{Zhou, S.-Qi}, \au{Sun, C.} \& \au{Xia,
  K.-Q.}} \yr{2009}  \at{Oscillations of the large-scale circulation in
  turbulent {R}ayleigh–{B}énard convection: the sloshing mode and its
  relationship with the torsional mode}.  \jt{J. Fluid. Mech.}  \bvol{630},
  \pg{367–390}.

\bibitem[Zürner {\em et~al.\/}(2019)Zürner, Schindler, Vogt, Eckert \&
  Schumacher]{Zuerner2019}
{\sc \au{Zürner, T.}, \au{Schindler, F.}, \au{Vogt, T.}, \au{Eckert, S.} \&
  \au{Schumacher, J.}} \yr{2019}  \at{Combined measurement of velocity and
  temperature in liquid metal convection}.  \jt{J. Fluid. Mech.}  \bvol{876},
  \pg{1108–1128}.

\end{thebibliography}

\end{document}